**Photomechanical coupling in photoactive nematic elastomers**

Ruobing Bai, Kaushik Bhattacharya*
Division of Engineering and Applied Science, California Institute of Technology, Pasadena, California 91125, USA
* Corresponding author: bhatta@caltech.edu

**Abstract**

Photoactive nematic elastomers are soft rubbery solids that undergo deformation when illuminated. They are made by incorporating photoactive molecules like azobenzene into nematic liquid crystal elastomers. Since its initial demonstration in 2001, it has received increasing interest with many recent studies of periodic and buckling behavior. However, theoretical models developed have focused on describing specific deformation modes (e.g., beam bending and uniaxial contraction) in the absence of mechanical loads, with only limited attention to the interplay between mechanical stress and light-induced deformation. This paper explores photomechanical coupling in a photoactive nematic elastomer under both light illumination and mechanical stress. We begin with a continuum framework built on the free energy developed by Corbett and Warner (*Phys. Rev. Lett.* 2006). Mechanical stress leads to nematic alignment parallel to a uniaxial tensile stress. In the absence of mechanical stress, in the photo-stationary state where the system reaches equilibrium, the nematic director tends to align perpendicular to the polarization of a linearly polarized light. However, sufficient illumination can destroy nematic order through a first-order nematic-isotropic phase transition which is accompanied by a snap through deformation. Combined illumination and mechanical stress can lead to an exchange of stability accompanied by stripe domains. Finally, the stress-intensity phase diagram shows a critical point that may be of interest for energy conversion.

**Keywords:** *liquid crystal elastomer, photochemistry, actuation, phase transformation, critical point*

## 1. Introduction

Photoactive nematic elastomers are soft rubbery solids that undergo deformation when illuminated. They are made by incorporating photoactive molecules like azobenzene into nematic liquid crystal elastomers. Since its initial demonstration by Finkelmann et al. (2001), it has received increasing interest with many recent studies of periodic and buckling behavior. Hogan et al. (2002) conducted experiments on a range of different photoactive nematic elastomers. Cviklinski et al. (2002) developed a photomechanical transducer that transforms light to mechanical force. Yu et al. (2003) showed clear evidence of photochemistry-induced actuation through directed bending of nematic films by polarized light. Recent reviews on this subject include Kuenstler and Hayward (2019), Mahimwalla et al. (2012), Pang et al. (2019), Ube and Ikeda (2014), and White (2018). Photomechanical actuation enables various functionalities including high-frequency beam vibration (White et al., 2008), on-demand shape morphing (Ahn et al., 2016), motors (Yamada et al., 2008), swimming robots (Camacho-Lopez et al., 2004), crawling robots (Gelebart et al., 2017), and rolling robots (Wie et al., 2016).

Light actuation of liquid crystal polymers and elastomers through photochemistry is attractive for several reasons. The actuation is untethered, fueled wirelessly by photons at the speed of light. Compared to photothermal actuation, photochemistry does not require additional light-absorbing particles that are usually immiscible in the polymer network. The variety in the chemistry of photochromophore, liquid crystal mesogen, and polymer network greatly expands the material space for desired working conditions (Kim et al., 2014). Other than light, photochemical actuation does not rely on special ambient environment such as large temperature change, pH, humidity, electric, or magnetic field. The actuation is mostly reversible and repeatable due to the reversibility of most photoisomerizations (Cviklinski et al., 2002). The actuation can be further designed with fast speed (White et al., 2008) or large work output (Dong et al., 2019). High tunability and accurate control can potentially be encoded in the light via its wavefront shape, wavelength, polarization, and intensity. Taking advantage of lasers with spot size of micrometers, the actuation forms an excellent candidate for driving micro-robots (Zeng et al., 2018).

Theoretical models have been developed for photochemistry induced deformation in liquid crystal elastomers (LCEs). However, most of them have been focused on describing specific actuation modes (e.g., beam bending and uniaxial contraction). In their original paper, Finkelmann et al. (2001) studied a nematic LCE embedded with azobenzene as photochromophore. They showed the reduction of nematic order parameter due to the *trans*-*cis* isomerization of azobenzene, and accounted the effect of this isomerization as an equivalent increase of temperature. The light actuation is subsequently treated using a 1D thermomechanical model, where the effective temperature increase depends on the intensity of light. Many more continuum theories were developed based on this analogy of temperature effect, essentially treating the photomechanical deformation as an effective thermal expansion (Hogan et al., 2002; Jin et al., 2010a; Knežević et al., 2013; Lin et al., 2012; Liu and Onck, 2017).

Another group of theoretical models focus on bending, due to the limited penetration depth of light into a LCE in most applications. Among them, Warner and Mahadevan (2004) derived a scaling analysis of photo-induced bending of beams, plates, and films. Corbett and Warner (2007) modeled photo-induced bending due to a nonlinear absorptive effect. Dunn (2007) studied the effect of polarization on bending of LCE films.

While these models predict the material response under specified loading conditions of interest and largely study free deformation with no applied mechanical loads, they do not describe detailed multiphysical interaction in the material during actuation. This lack of microscopic details hinders the opportunity of discovering new actuation modes, or exploring material behaviors under more complex conditions.

Besides these models of specific actuation modes, theories combining statistical mechanics and continuum mechanics were also developed. Corbett and Warner (2006, 2008) modeled light-induced deformation of a stress-free polydomain nematic LCE by investigating the coupling between the polymer network elasticity and mesogen mixture. Despite this effort, an important question has yet been answered: how does mechanical stress affect the light-induced deformation in a LCE? A full coupling between light illumination and mechanical stress, as well as the resulting photomechanical response of the material, is still largely unexplored.

This paper explores photomechanical coupling in a photoactive LCE under both light illumination and mechanical stress. We begin with a continuum framework in Section 2 built on the free energy developed by Corbett and Warner (2006). We focus on the photo-stationary state where the system reaches equilibrium. Through energy minimization with homogeneous deformation, we study ground states in Section 3, and show that the nematic director tends to align perpendicular to the polarization of a linearly polarized light in Section 4. Section 5 studies the competition between illumination and stress since the director aligns parallel to any applied uniaxial stress. We demonstrate such photo-alignment and mechano-alignment, and the induced large deformation in a sheet of nematic LCE under both light illumination and mechanical stress. The transition between these two modes induces a nonzero shear strain, giving rise to formation of stripe domains in the material. In Section 6, we show the first-order nematic-isotropic phase transformation as the light intensity increases, and a corresponding snap-through instability in the deformation of the material. The phase diagram under various light intensity and mechanical stress is plotted, and a critical point is identified. The implications of all these findings are discussed in Section 7.

## 2. Continuum theory of photoactive nematic liquid crystal elastomers

### 2.1. Microscopic picture

The microscopic process of photochemistry induced large deformation is illustrated in Fig. 1a. Here we focus on a monodomain main-chain liquid crystal elastomer where the liquid crystal mesogens are linked as parts of the main-chain backbone of a crosslinked polymer network. Photochromophores such as azobenzene molecules are embedded in the elastomer, either by forming bonds with the network, or by simply mixing with the network. At a temperature lower than the nematic-isotropic transition temperature, the mesogens stay in the nematic state, aligning along a direction due to steric effect or weak intermolecular interaction such as dipole-dipole. The average aligned direction is defined by the nematic director $\boldsymbol{n}$. The degree of alignment is defined by the nematic order parameter $Q$.

When illuminated with a light of certain wavelength, azobenzene absorbs photons, and isomerize from the *trans*-state to the *cis*-state. The bent shape of *cis*-state reduces the nematic order in the elastomer, and introduces a macroscopic deformation of the elastomer network.

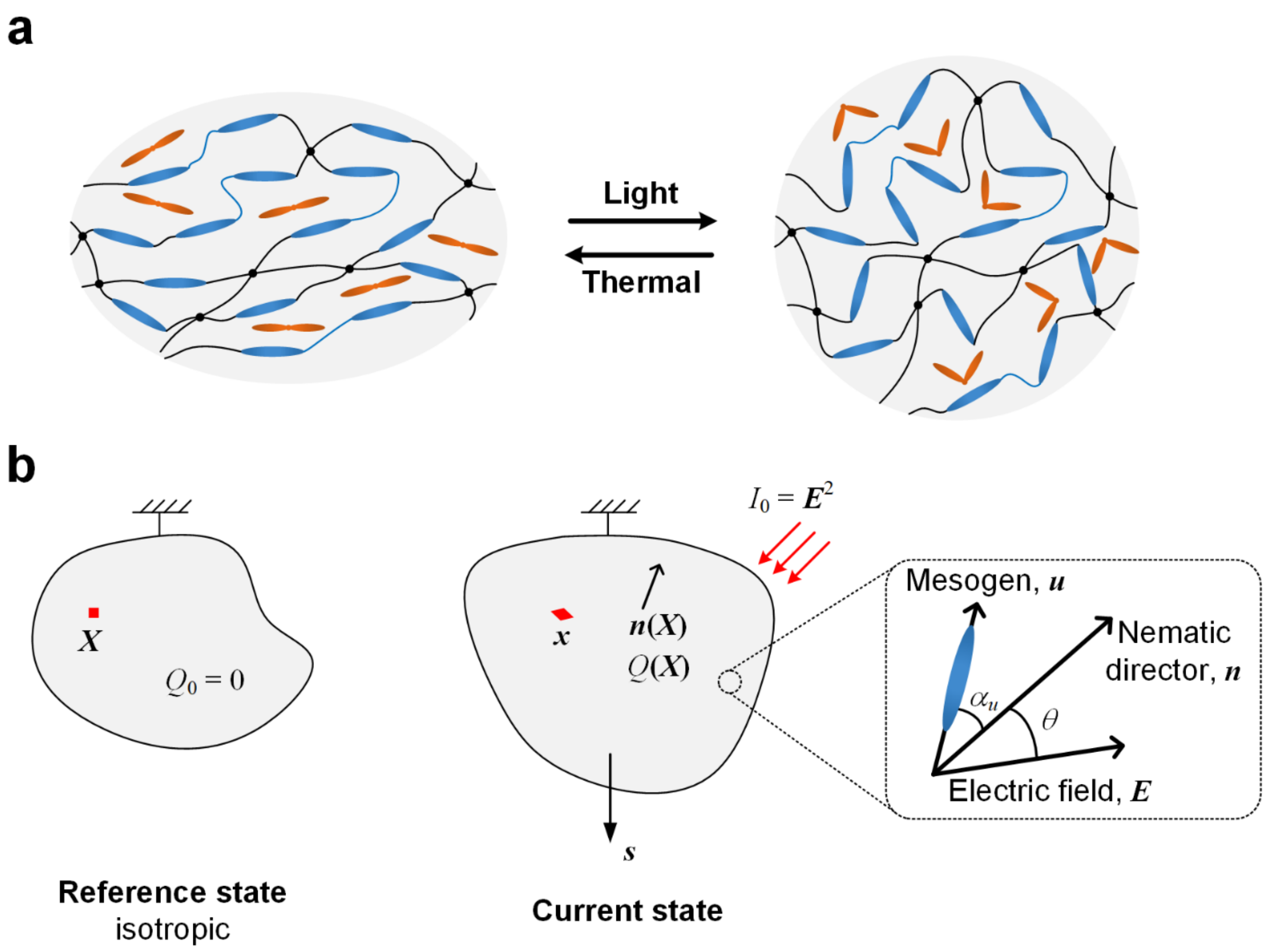

**Fig. 1.** Formulation of the continuum theory. **(a)** Microscopic process of photochemistry induced large deformation in a main-chain nematic liquid crystal elastomer (blue rods: non-photoactive mesogens; black lines: polymer chains; black dots: crosslinks; orange rods: photochromophores). Under illumination with light of a certain wavelength, the photochromophores such as azobenzene isomerize, reduce the nematic order in the elastomer, and introduce a macroscopic deformation of the elastomer network. **(b)** Continuum framework. The undeformed isotropic phase is taken as the reference state. The nematic phase under a nominal stress field $\boldsymbol{s}$ and a linearly polarized light with intensity $I_0$ is taken as the current state. $\boldsymbol{X}$ represents the coordinate of a material particle in the reference state, and $\boldsymbol{x}(\boldsymbol{X})$ represents the same material particle in the current state. In the current state, the nematic director is $\boldsymbol{n}(\boldsymbol{X})$ and the nematic order parameter is $Q(\boldsymbol{X})$.

### 2.2. Continuum framework

We begin with a continuum framework following a free energy developed by Corbett and Warner (2006). We focus on the photo-stationary state where the system reaches equilibrium. We neglect the radiation pressure from the light, which is typically smaller than the modulus of the elastomer by orders of magnitude (Bai and Suo, 2015).

We take the undeformed isotropic phase as the reference state and label particles with their position $\boldsymbol{X}$ in the reference state. The current position of the particle is denoted $\boldsymbol{x}(\boldsymbol{X})$, the order parameter is denoted $Q(\boldsymbol{X})$, and the director is denoted $\boldsymbol{n}(\boldsymbol{X})$. The order parameter is zero in the isotropic phase, and non-zero in the nematic phase. The director is indeterminate in the isotropic phase. The deformation gradient is defined as

$$F_{iK} = \frac{\partial x_i}{\partial X_K}. \tag{2.1}$$

We denote $W$ as the density of Helmholtz free energy stored in the deformed material. Following Corbett and Warner (2006, 2008), we write the Helmholtz free energy $W$ as

$$W = W_{\mathrm{e}} + W_{\mathrm{lc}}, \tag{2.2}$$

where $W_{\mathrm{e}}$ is the entropic elastic energy of the polymer network, and $W_{\mathrm{lc}}$ is the free energy due to the mixing of nematic mesogens. The free energy of mixing between mesogens and polymer chains is neglected, assuming no migration of mesogens during the deformation. In this model, $W_{\mathrm{lc}}$ depends on the concentration of the isomerized *cis*-chromophores in the equilibrium state. The concentration of *cis*-chromophores is further determined by the photoreaction, which depends on the light intensity and the coupling angle between the nematic director and light polarization.

In the current model, we also neglect the Frank elasticity induced by the distortion of mesogen alignment (de Gennes and Prost, 1995; Frank, 1958). The Frank elastic constant $K$ and the modulus of the elastomer $\mu$ form a length scale $\sqrt{K/\mu} \sim 10^{-8}$ m, indicating a representative length of domain boundary such as stripe domains that can form in a LCE (Warner and Terentjev, 2003). The current model can be readily generalized to include the Frank elasticity if the detail of microstructures is of particular interest.

### 2.3. Entropic elastic energy of the polymer network

We take the anisotropic Gaussian chain model for the entropic elastic energy of the polymer network (Bladon et al., 1993; Warner and Terentjev, 2003)

$$W_{\mathrm{e}}\left(\boldsymbol{F}, Q, \boldsymbol{n}\right) = \frac{NkT}{2}\left[\mathrm{Tr}\left(\boldsymbol{F}^T \cdot \boldsymbol{l}^{-1} \cdot \boldsymbol{F}\right) + \log\left(\det \boldsymbol{l}\right)\right], \tag{2.3}$$

where $N$ is the number of polymer chains per unit volume, and $kT$ is the temperature in the unit of energy. $\boldsymbol{l}$ is the reduced shape tensor in the current state, expressed as

$$l_{ij} = (1-Q)\delta_{ij} + 3Qn_i n_j , \tag{2.4}$$

with its inverse

$$\left(l^{-1}\right)_{ij} = \frac{1}{1-Q}\delta_{ij} + \left(\frac{1}{1+2Q} - \frac{1}{1-Q}\right) n_i n_j . \tag{2.5}$$

Note that in the isotropic phase, $Q = 0$ and $\boldsymbol{l}$ is identity, and (2.3) recovers the neo-Hookean model.

**2.4. Maier-Saupe mean field free energy**

We take the modified Maier-Saupe mean field model for the free energy of the nematic mixture (see Appendix A for the derivation) (Corbett and Warner, 2006, 2008; Maier and Saupe, 1959)

$$W_{\mathrm{lc}}(Q,\boldsymbol{n}) = N_n kT\left(1-c(Q,\boldsymbol{n})\right)\left[g^{-1}(Q)Q - \log Z(Q) - \frac{1}{2}\left(1-c(Q,\boldsymbol{n})\right)\frac{J}{kT}Q^2\right], \tag{2.6}$$

where $N_n$ is the total number of mesogens (*trans*-, *cis*-, and non-photoactive) per unit volume, $c$ the fraction of *cis*-mesogens, $g^{-1}(Q)$ the trial field, $Z(Q)$ the partition function, and $J$ the average interaction between two mesogens in the unit of energy. The expressions of $g(Q)$ and $Z(Q)$ are

$$g(x) = -\frac{1}{2} - \frac{1}{2x} + \frac{1}{2x}\sqrt{\frac{3x}{2}}\frac{\exp(3x/2)}{\int_0^{\sqrt{3x/2}}\exp\left(y^2\right)\mathrm{d}y}, \tag{2.7}$$

$$Z(Q) = \frac{\exp\left[g^{-1}(Q)\right]}{1+g^{-1}(Q)(1+2Q)} . \tag{2.8}$$

In this model, the *trans*-mesogens and the non-photoactive mesogens are assumed to be statistically identical to form the nematic mixture. The *trans*-*cis* photo-isomerization has a forward reaction driven by light and a backward reaction driven by thermal relaxation. The model assumes a first-order reaction at the single-molecule level. That is, the reaction equilibrium of a chromophore molecule is independent of interaction from the surrounding molecules. As a result, the fraction of the *cis*-mesogen is expressed as (Appendix B)

$$c = c(Q,\theta) = f\frac{I\left[1+Q\left(3\cos^2\theta-1\right)\right]}{3+I\left[1+Q\left(3\cos^2\theta-1\right)\right]}, \tag{2.9}$$

where $I = \boldsymbol{E}^2\Gamma\tau$ is the dimensionless intensity of light, and $\theta$ is the angle between the nematic director $\boldsymbol{n}$ and the light polarization $\boldsymbol{E}$. The isomerized *cis*-mesogens reduce the nematic order, reflected by the factor of $(1-c)$ in Equation (2.6). Additional parameters may be introduced to account for the difference between photoactive and non-photoactive mesogens (Corbett and Warner, 2008), which is not included in the current model.

**2.5. Equilibrium**

We consider a material under homogeneous deformation $F_{iK}$ and nominal stress $s_{iK}$. The total potential energy of the system can be written as

$$\hat{W} = W(\boldsymbol{F},Q,\boldsymbol{n}) - s_{iK}F_{iK}. \tag{2.10}$$

where $-s_{iK}F_{iK}$ is the potential energy of the mechanical stress. With a prescribed stress, $\hat{W}$ is minimized in the equilibrium state, leading to

$$s_{iK}(\boldsymbol{F},Q,\boldsymbol{n}) = -p\left(\boldsymbol{F}^{-1}\right)^T + \frac{\partial W(\boldsymbol{F},Q,\boldsymbol{n})}{\partial F_{iK}}, \tag{2.11}$$

$$\frac{\partial W(\boldsymbol{F},Q,\boldsymbol{n})}{\partial Q} = 0, \tag{2.12}$$

$$\boldsymbol{n}\times\frac{\partial W(\boldsymbol{F},Q,\boldsymbol{n})}{\partial n_i} = 0. \tag{2.13}$$

subject to

$$\det\boldsymbol{F} = 1, \tag{2.14}$$

$$n_i n_i = 1. \tag{2.15}$$

In the above equation, we have assumed incompressibility of the elastomer, so that $p$ is the Lagrangian multiplier represented by a hydrostatic pressure, and will be determined by solving the boundary value problem.

### 2.6. Dimensionless groups

For the rest of this paper, we use dimensionless groups and their corresponding values listed in Table 1.

**Table 1.** Dimensionless groups and their values

| Normalized term | Dimensionless group | Value |
|---|---|---|
| Shear modulus | $\tilde{\mu} = (NkT)/(N_n kT)$ | 0.05 |
| Stress | $\tilde{s} = s/\mu = s/(NkT)$ | Variable values |
| Light intensity | $I = \boldsymbol{E}^2 \Gamma \tau$ | Variable values |
| Interaction between mesogens | $\tilde{J} = J/kT$ | 5 |
| Fraction of photoactive mesogens | $f$ | 1/6 |

## 3. Ground state under no mechanical stress or light illumination

### 3.1. General form of the ground state deformation

We first investigate the ground state deformation under no mechanical stress or light illumination. In this case, $c = 0$ and $s_{iK} = 0$. From (2.6), the free energy of the nematic mixture is

$$W_{\mathrm{lc}} = N_n kT \left[ g^{-1}(Q)Q - \log Z(Q) - \frac{1}{2}\frac{J}{kT}Q^2 \right], \tag{3.1}$$

which is independent of the deformation gradient ***F***. Assuming a homogeneous deformation and minimizing the total free energy $W = W_{\mathrm{e}} + W_{\mathrm{lc}}$ with respect to ***F***, we obtain the general ground state deformation gradient

$$\boldsymbol{F} = \frac{1}{(\det \boldsymbol{l})^{1/6}} \boldsymbol{R}_2 \cdot \boldsymbol{l}^{\frac{1}{2}} \cdot \boldsymbol{R}_1, \tag{3.2}$$

where $\boldsymbol{R}_1$ and $\boldsymbol{R}_2$ are two arbitrary rotation tensors applied to the material in the reference state and current state, respectively. Note that there are infinite ground states in equilibrium. Each ground state deformation can be understood by decomposing the deformation gradient into steps as indicated in Fig. 2.

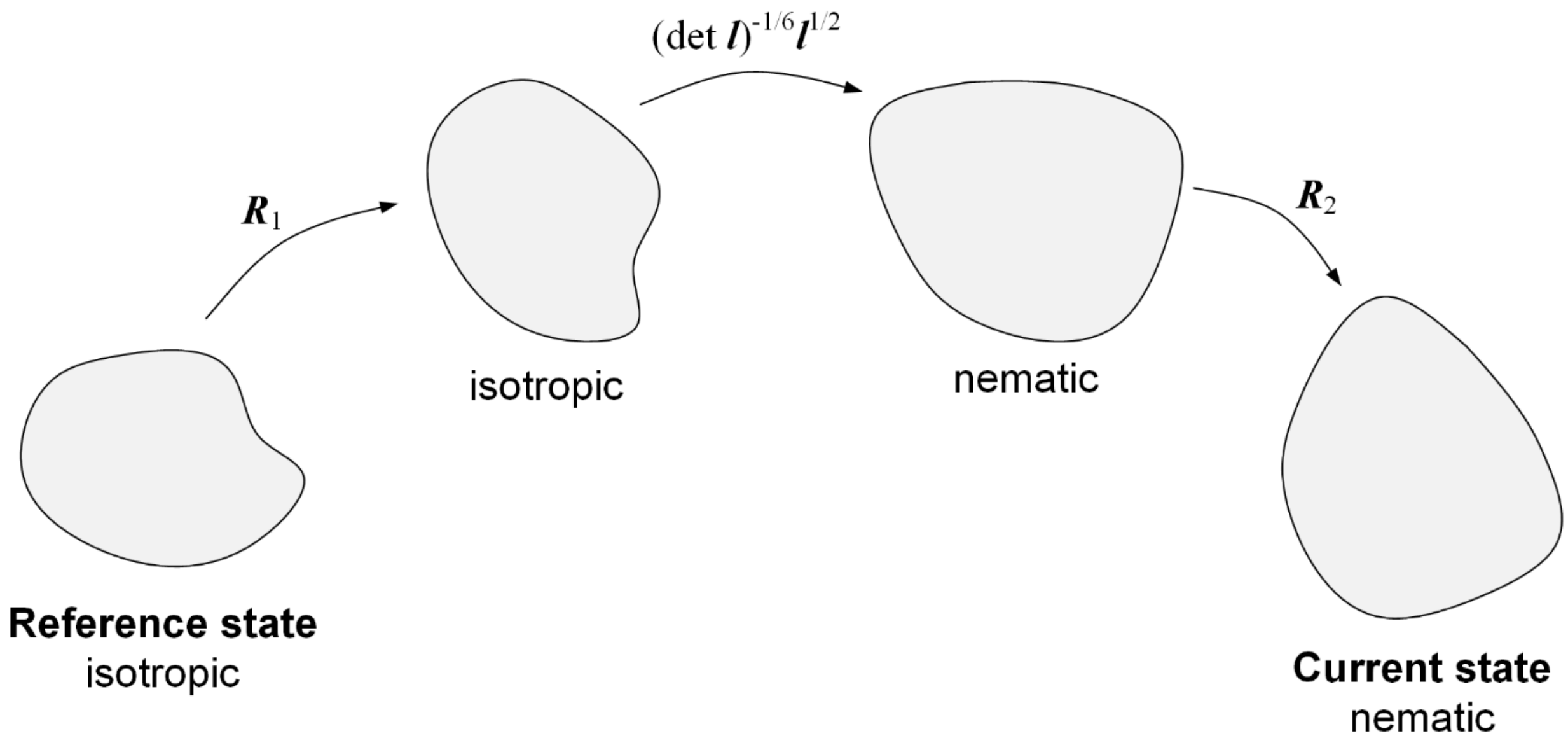

**Fig. 2**. Decomposition of a ground state deformation gradient (denoted as the *Current state* in the figure).

Without losing generality, we can take $\boldsymbol{l}$ to be diagonal in a rectangular frame,

$$\boldsymbol{l} = \begin{bmatrix} 1+2Q & & \\ & 1-Q & \\ & & 1-Q \end{bmatrix}. \tag{3.3}$$

The deformation gradient written in its principal coordinates is then

$$\frac{1}{\left(\det \boldsymbol{l}\right)^{1/6}} \boldsymbol{l}^{\frac{1}{2}} = \begin{bmatrix} r^{1/3} & & \\ & r^{-1/6} & \\ & & r^{-1/6} \end{bmatrix}. \tag{3.4}$$

where $r = \left(1+2Q\right)/\left(1-Q\right)$. Because we have taken the isotropic phase as the reference state, this deformation gradient corresponds to the spontaneous deformation of the LCE in the nematic phase due to the isotropic-nematic transition. The first principal stretch, $r^{1/3}$, corresponds to the elongation along the nematic director.

### 3.2. Soft elastic deformation and stripe domains

In addition to the infinite ground states, we can construct additional effective ground states by stripe domains (Bhattacharya, 2003; DeSimone and Dolzmann, 2002; Verwey et al., 1996). Let

$$\bar{\boldsymbol{F}} = \begin{bmatrix} \bar{F}_{11} & & \\ & \bar{F}_{22} & \\ & & \bar{F}_{33} \end{bmatrix}, \tag{3.5}$$

for $r^{-1/6} < \bar{F}_{11}, \bar{F}_{33} < r^{1/3}$. By (3.2), this deformation gradient is not a ground state deformation. However, we now show that it can be achieved by alternating layers of ground states $\boldsymbol{F}^{+}$ and $\boldsymbol{F}^{-}$ as shown in Fig. 3. Let

$$\boldsymbol{F}^{\pm} = \frac{1}{(\det \boldsymbol{l})^{1/6}} \boldsymbol{R}_2^{\pm} \cdot \boldsymbol{l}^{\frac{1}{2}} \cdot \boldsymbol{R}_1^{\pm}, \tag{3.6}$$

where $\boldsymbol{l}$ is given in (3.3), and the two planar rotation tensors are

$$\boldsymbol{R}_1^{\pm} = \begin{bmatrix} \cos\psi & 0 & \mp\sin\psi \\ 0 & 1 & 0 \\ \pm\sin\psi & 0 & \cos\psi \end{bmatrix}, \ \boldsymbol{R}_2^{\pm} = \begin{bmatrix} \cos\phi & 0 & \mp\sin\phi \\ 0 & 1 & 0 \\ \pm\sin\phi & 0 & \cos\phi \end{bmatrix}, \tag{3.7}$$

with rotation angles $\pm\psi$ and $\pm\phi$. The nematic directors corresponding to $\boldsymbol{F}^{\pm}$ are $\boldsymbol{n}^{\pm} = (\cos\phi, 0, \pm\sin\phi)^T$ after rotation. Clearly ( $\boldsymbol{F}^{\pm}$, $\boldsymbol{n}^{\pm}$ ) are ground states.

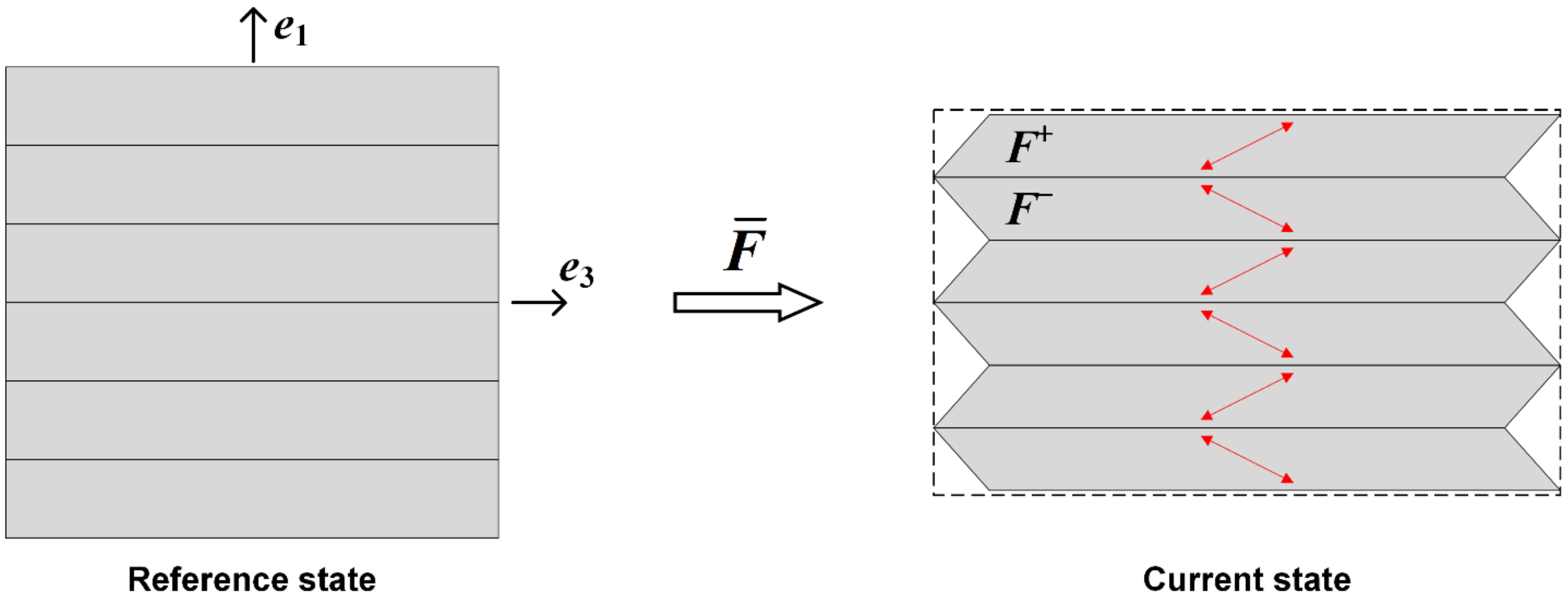

**Fig. 3.** Formation of stripe domains with an alternating layers of ground states $\boldsymbol{F}^{+}$ and $\boldsymbol{F}^{-}$. The average deformation gradient $\bar{\boldsymbol{F}}$ has no shear strain. Red arrows indicate the nematic directors in the stripe domains, $\boldsymbol{n}^{\pm} = (\cos\phi, 0, \pm\sin\phi)^T$.

For two neighboring stripe domains in Fig. 3, we can always find a pair of angles $(+\psi, +\phi)$ and

$\left(-\psi,-\phi\right)$, such that $F_{13}^{+}=F_{13}^{-}=0$. This is satisfied by a single equation

$$r^{1/3}\sin\psi\cos\phi+r^{-1/6}\cos\psi\sin\phi=0\,. \tag{3.8}$$

Under this condition, note that

$$\boldsymbol{F}^{+}-\boldsymbol{F}^{-}=\boldsymbol{a}\otimes\boldsymbol{N} \tag{3.9}$$

is satisfied with $\boldsymbol{a}=\left(0,0,2F_{31}^{+}\right)^{T}$ and $\boldsymbol{N}=\left(1,0,0\right)^{T}$, where $F_{31}^{+}=r^{-1/6}\sin\psi\cos\phi+r^{1/3}\cos\psi\sin\phi$. (3.9) is the Hadamard's compatibility condition that is required at the interface between two domains with compatible deformation gradient $\boldsymbol{F}^{+}$ and $\boldsymbol{F}^{-}$ (Bhattacharya, 2003). The average deformation gradient of the bulk material is then

$$\left\langle\boldsymbol{F}\right\rangle=\alpha\boldsymbol{F}^{+}+\left(1-\alpha\right)\boldsymbol{F}^{-}=\bar{\boldsymbol{F}}\,, \tag{3.10}$$

with $\alpha$ = 1/2. The average free energy is

$$\left\langle W\right\rangle=\alpha W\left(\boldsymbol{F}^{+},Q,\boldsymbol{n}^{+}\right)+\left(1-\alpha\right)W\left(\boldsymbol{F}^{-},Q,\boldsymbol{n}^{-}\right), \tag{3.11}$$

which is equal to the free energy of the ground state, and lower than that considering homogeneous deformation, $W\left(\bar{\boldsymbol{F}},Q,\boldsymbol{n}\right)$. Within each domain, $F_{13}=0$ and $F_{31}\neq 0$, but the average deformation has zero shear strain, $\bar{F}_{13}=\bar{F}_{31}=0$ so that the overall deformation gradient has the form (3.5).

This infinite number of ground states and formation of stripe domains are usually shown in experiments by the "soft elastic" behavior of a LCE (Küupfer and Finkelmann, 1994; Olmsted, 1994; Warner et al., 1994). In such an experiment, a sheet of a monodomain nematic LCE is stretched perpendicular to its initial nematic director. During stretch, the nematic director rotates towards the stretching direction, with the recorded stress nearly zero. After the nematic director aligns with the stretching direction, the stress starts to increase, and the LCE behaves as a normal elastic rubber. This soft elastic behavior is less significant when a LCE is crosslinked in the nematic phase rather than isotropic phase, possibly due to the memory of the initial state stored in the crosslinked network (Biggins et al., 2012). As will be shown in Section 5, the formation of stripe domains also exists when the material is under both light illumination and mechanical stress.

## 4. Photo-alignment under no mechanical stress

We next investigate the free energy landscape and equilibrium state of a 2D nematic sheet under light illumination but without mechanical stress. In the current state, a linearly polarized light is illuminated perpendicular onto the surface of the sheet (Fig. 4). The equilibrium angle between the nematic director and polarization is $\theta$. We assume that the sheet is much thinner than the decay length of light in the LCE, such that the light intensity is homogeneous and prescribed as the dimensionless $I$.

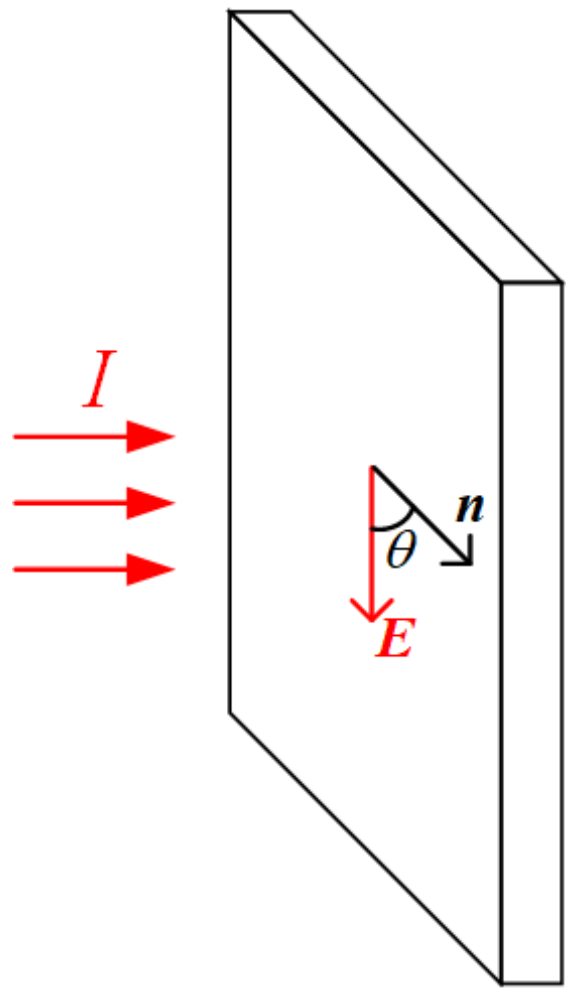

**Fig. 4**. A linearly polarized light is illuminated perpendicular onto the surface of a thin sheet of nematic elastomer. In the equilibrium state, the angle between the nematic director and polarization is $\theta$.

With no applied stress, the total potential energy in the material is

$$W(\boldsymbol{F},Q,\theta)=W_{\mathrm{e}}(\boldsymbol{F},Q,\theta)+W_{\mathrm{lc}}(Q,\theta), \tag{4.1}$$

where the nematic director is represented by the angle $\theta$. We minimize $W(\boldsymbol{F},Q,\theta)$ to find the equilibrium $\boldsymbol{F}$, $Q$, and $\theta$. We note that $W_{\mathrm{lc}}(Q,\theta)$ does not depend on the deformation gradient $\boldsymbol{F}$. As a result, minimizing $W$ with respect to $\boldsymbol{F}$ readily leads to the general ground state deformation shown in (3.2), even under a finite light intensity. We express the effective total energy as $\min_{\boldsymbol{F}} W(\boldsymbol{F},Q,\theta)=\tilde{W}(Q,\theta)$, and plot its contour in the polar coordinates $(Q,\theta)$ under different light intensity $I$ in Fig. 5. The polarization is set along $\theta=0$.

Together with each energy contour, we also plot the energy curves at $\theta = 0$ and $\theta = \pi / 2$, as well as the minimum energy $\tilde{W}\left(Q = Q_{\min}, \theta\right)$ along the azimuthal direction $\theta$.

The global minimum of the energy landscape indicates a photo-alignment under light illumination. The contour has a 4-fold symmetry in $\theta$ $\left(\boldsymbol{n} \leftrightarrow \text{-}\boldsymbol{n}, \theta \leftrightarrow -\theta\right)$, so we focus on $\theta \in \left[0, \pi / 2\right]$. With no light illumination $I$ = 0 (Fig. 5a), the energy contour is axisymmetric, indicating no preferred alignment of the nematic director. The energy reaches minimum at a finite order parameter $Q = 0.61$, indicating a nematic phase. As the light intensity increases ($I$ = 0.5, Fig. 5b), the minimum at $\theta = 0$ becomes a saddle point, and the global minimum moves to $\theta = \pi / 2$. The minimizing order parameter remains finite (though smaller than 0.61), so the LCE remains in the nematic phase. Recall that the light polarization is fixed at $\theta = 0$. Therefore, a linearly polarized light with finite intensity aligns the nematic director perpendicular to the polarization. Furthermore, since the nematic director under such photo-alignment is fixed at $\theta = \pm\pi / 2$, the material loses the additional freedom of rotation to form stripe domains.

As the light intensity increases further to $I$ = 8, the energy landscape along $\theta = \pi / 2$ shows a double-well feature, with a global minimum at $Q = 0$ and a local minimum at a finite $Q$ (Fig. 5c). This indicates a first-order, nematic-isotropic phase transformation, similar to the phase transformation induced by temperature increase (Warner and Terentjev, 2003). More details of this first-order phase transformation and its coupling with the mechanical stress will be discussed in Section 5 and 6.

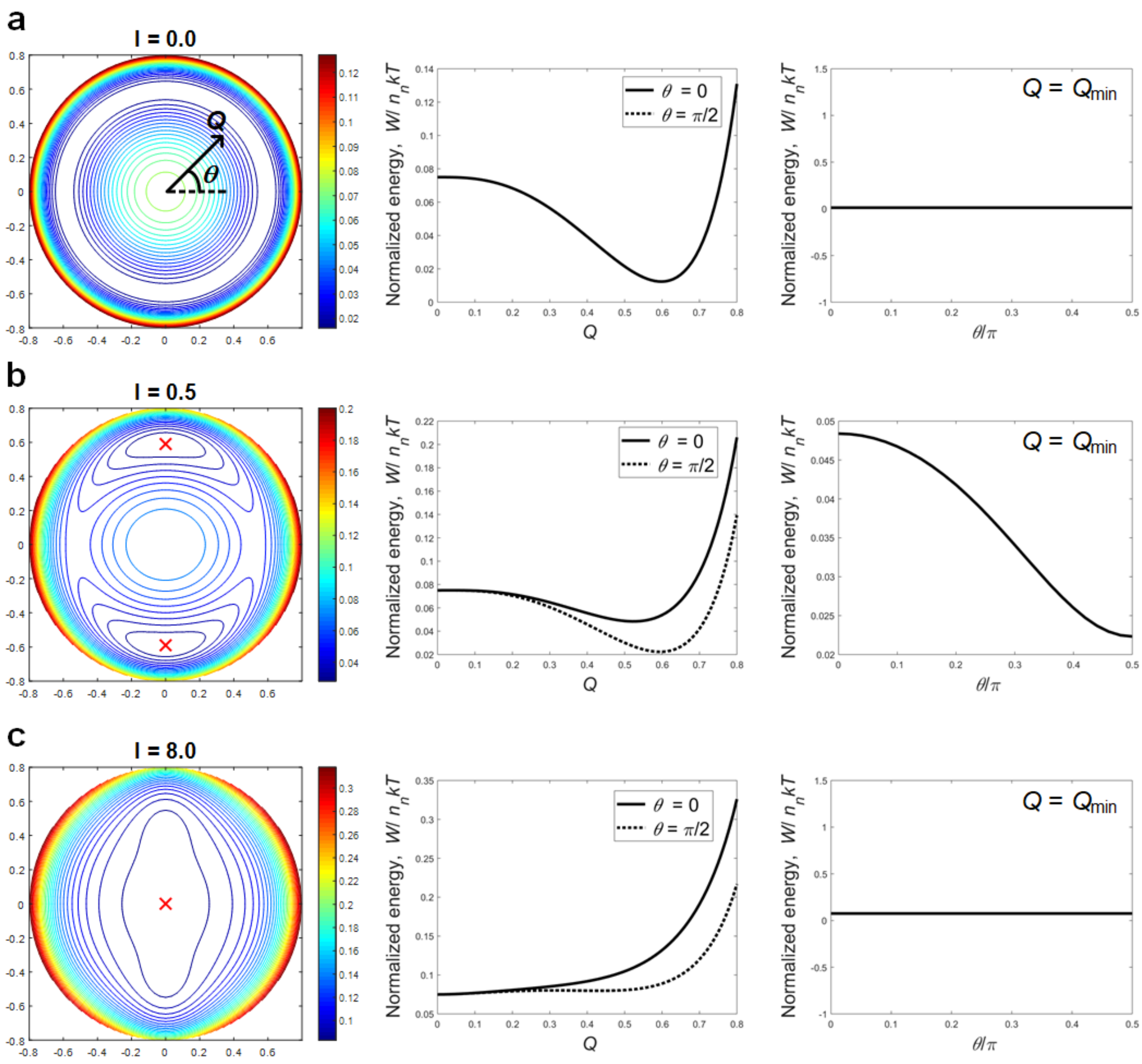

**Fig. 5**. Contours of total energy under increasing light intensity and no mechanical stress. $\theta$ represents the angle between the nematic director and polarization. The red cross marks the global energy minimum. Together with each energy contour are the energy curves at $\theta = 0$ and $\theta = \pi / 2$, as well as the minimum energy $\tilde{W}\left(Q = Q_{\min}, \theta\right)$ along the azimuthal direction $\theta$. **(a)** When $I = 0$, the energy reaches minimum at a finite $Q$ and arbitrary $\theta$. **(b)** When $I = 0.5$, the energy reaches global minimum at $\theta = \pm\pi / 2$, while the minimum along $\theta = 0$ becomes a saddle point. **(c)** When $I = 8$, the energy reaches global minimum at $Q = 0$ and local minimum at a finite $Q$ along $\theta = \pm\pi / 2$.

Photo-alignment of liquid crystal photochromophores such as azobenzene under polarized light has been extensively observed and studied in solutions and polymer melts (Ichimura, 2000; Ikeda, 2003). An explanation for this phenomenon is the cyclic forward and backward reactions in the photo-stationary state (reaction equilibrium). The photoactive azobenzene reaches strongest coupling with light when its nematic director is parallel to the polarization, but weakest coupling when the nematic director is perpendicular. In the photo-stationary state, the cyclic forward and backward reactions make a chromophore undergo the *trans*-*cis*-*trans* process and so on. As a result, the remaining *trans*-molecules in the equilibrium state will be aligned nearly perpendicular to the polarization, so that they have weakest coupling with light.

Deformation of a crosslinked liquid crystal network induced by photo-alignment was also reported in experiments (Pang et al., 2019; Ube and Ikeda, 2014). Ideally, such a reorientation of nematic director can be acommodated by the soft elastic deformation of the polymer network. However, the capability of this soft elastic response under large deformation highly depends on the crosslinking history of the liquid crystal polymer network, as has been shown both by experiment (Urayama et al., 2009) and theoretical analysis (Biggins et al., 2009, 2012). In general, soft elastic deformation can be attained in isotropic genesis elastomers (i.e., crosslinked in the isotropic phase) but is hardly attained in nematic genesis elastomers (i.e., crosslinked in the nematic phase). In the latter case, we expect that the reorientation of nematic director by light will be difficult, and large deformation will be mainly induced by the *trans*-*cis* photo-isomerization of chromophores or the subsequent nematic-isotropic phase transition with increasing light intensity. These two effects will be shown in Section 6.

A particular material system has other factors affecting the photo-alignment, the *trans*-*cis* photo-isomerization, and the nematic-isotropic phase transition. These factors include temperature (compared to the nematic-isotropic transition temperature), type and concentration of photochromophores in the system, type of the liquid crystal polymer (e.g., main-chain or side-chain), and the state of mechanical stress. All these factors deserve further investigation in experiments.

The current theoretical model can be extended to incorporate the crosslinking history by adding additional free energy rising from non-ideality of the anisotropic Gaussian chain model in (2.3) (Biggins et al., 2009, 2012; Warner and Terentjev, 2003), and to incorporate the Frank elasticity due to the distortion of nematic director (de Gennes and Prost, 1995; Frank, 1958), which may set an energy barrier for photo-alignment. The dominant mechanism for the material response in different scenarios can be explored through a parameter study (e.g. dimensionless groups in Table 1) of different factors listed above. In the next two sections, we will focus on mechanical stress as one particular factor, to study its effect on the photo-alignment, the *trans*-*cis* photo-isomerization, and the nematic-isotropic phase transition.

## 5. Photo-alignment vs. mechano-alignment

We now investigate the free energy landscape and equilibrium state of the 2D nematic sheet under both light illumination and mechanical stress. To understand the photomechanical interaction, we illuminate the sheet with a light polarization parallel to the applied nominal stress $s$, as shown in Fig. 6.

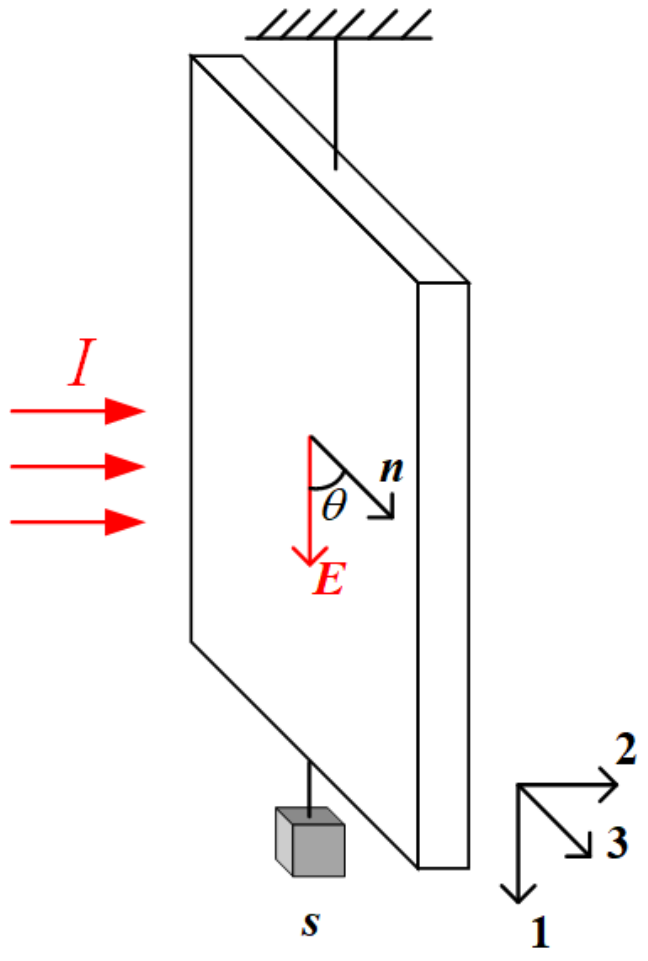

**Fig. 6**. A linearly polarized light is illuminated perpendicular onto the surface of a thin sheet of nematic elastomer. The elastomer is subject to a uniaxial nominal stress $s$ parallel to the polarization. In the equilibrium state, the angle between the nematic director and polarization is $\theta$.

The total potential energy of the system is

$$\hat{W}(\boldsymbol{F},Q,\theta)=W_{\mathrm{e}}(\boldsymbol{F},Q,\theta)+W_{\mathrm{lc}}(Q,\theta)-s\lambda_{11}, \tag{5.1}$$

where $-s\lambda_{11}$ is the potential energy of the mechanical stress, and $\lambda_{iK}$ are stretch components of the deformation gradient. Substituting in the general deformation gradient in 2D, we obtain

$$\begin{aligned}W_{\mathrm{e}}(\boldsymbol{F},Q,\theta)=&\frac{\mu}{2}\left(\lambda_{11}^2+\lambda_{13}^2\right)\left(\frac{\sin^2\theta}{1-Q}+\frac{\cos^2\theta}{1+2Q}\right)+\frac{\mu}{2}\left(\lambda_{31}^2+\lambda_{33}^2\right)\left(\frac{\cos^2\theta}{1-Q}+\frac{\sin^2\theta}{1+2Q}\right)+\frac{\mu}{2}\frac{\lambda_{22}^2}{1-Q}-\\&\frac{\mu}{2}\left(\frac{1}{1-Q}-\frac{1}{1+2Q}\right)\left(\lambda_{11}\lambda_{31}+\lambda_{13}\lambda_{33}\right)\sin 2\theta+\frac{\mu}{2}\log\left[(1-Q)^2\left(1-Q+3Q\cos^2\theta\right)\right].\end{aligned} \tag{5.2}$$

To plot the contour of the total energy in the polar coordinates $(Q,\theta)$ for each prescribed $s$ and $I$,

we minimize $\hat{W}\left(\boldsymbol{F},Q,\theta\right)$ with respect to $\boldsymbol{F}$ and numerically solve for the stretches. Under a fixed stress $s = 0.25\mu$, Fig. 7 shows the energy contours under different light intensity $I$, and Fig. 8 shows the stretches and nematic director angle $\theta$ as functions of $I$.

Due to the 4-fold symmetry in $\theta$ $\left(\boldsymbol{n}\leftrightarrow\boldsymbol{\text{-}n},\theta\leftrightarrow-\theta\right)$, there are generally 4 minima in the energy contour, so we focus on $\theta\in\left[0,\pi/2\right]$. When $I = 0$, the energy reaches global minimum at $\theta = 0$ (Fig. 7a), indicating that the uniaxial stress aligns the nematic director along its direction, which we denote as *mechano-alignment*. As $I$ increases to 0.13 (Fig. 7b), the photo-alignment and mechano-alignment become comparable, and the nematic director starts to rotate towards $\theta = \pi/2$ upon further increase of $I$. Within this transition region of rotation, e.g., $I = 0.18$ in Fig. 7c, the global minimum is located at $0 < \theta < \pi/2$, while the minima at $\theta = 0$ and $\pi/2$ become two saddle points. At $I = 0.25$ (Fig. 7d), the nematic director reaches $\theta = \pi/2$, and photo-alignment dominates. The energy reaches global minimum at $\theta = \pm\pi/2$.

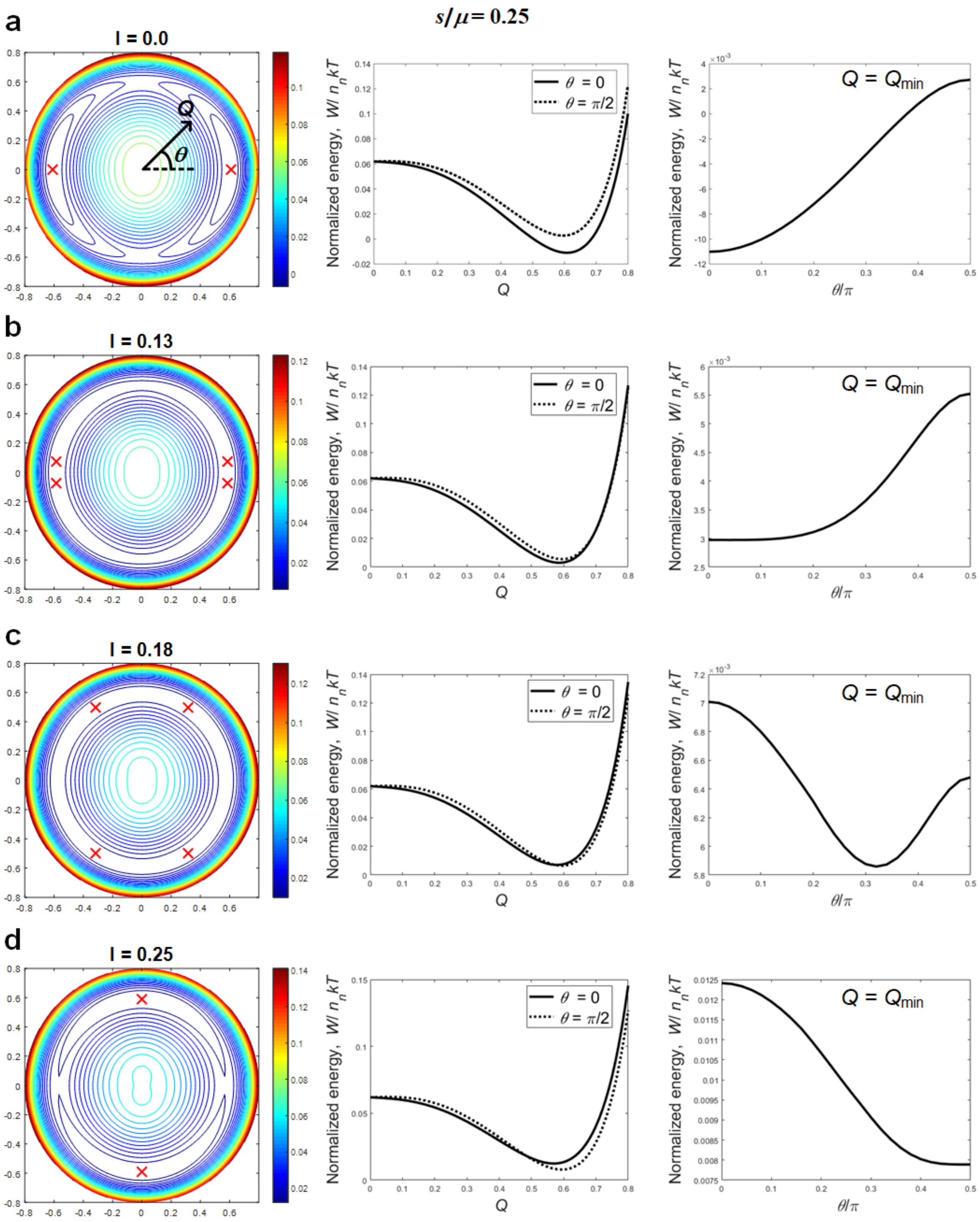

**Fig. 7**. The total energy contour in the polar coordinates $(Q,\theta)$ under a fixed stress $s = 0.25\mu$ and increasing light intensity $I$. The red cross marks the global energy minimum. $\theta$ describes the angle between the light polarization and nematic director. The directions of both the uniaxial stress and polarization are fixed at $\theta = 0$. Together with each energy contour are the energy curves at $\theta = 0$ and $\theta = \pi / 2$, as well as the minimum energy $\tilde{W}(Q = Q_{\min}, \theta)$ along the azimuthal direction $\theta$. **(a)** When $I = 0$, the energy reaches global

minimum at $\theta = 0$. **(b)** When $I = 0.13$, the director starts to rotate towards $\theta = \pm\pi/2$. **(c)** When $I = 0.18$, the global minimum is located at $0 < \theta < \pi/2$, indicating the transition from mechano-alignment to photo-alignment. **(d)** When $I = 0.25$, the nematic director reaches $\theta = \pi/2$, and photo-alignment dominates. The energy reaches global minimum at $\theta = \pm\pi/2$.

Fig. 8 plots the stretches and nematic director angle $\theta$ under the increasing light intensity. The stretch $\lambda_{11}$ is large when the light intensity $I$ is small, as the finite stress elongates the material and aligns the nematic director along the direction $\boldsymbol{e}_1$. Transition from mechano- to photo-alignment takes place at an intermediate $I$, where $\lambda_{11}$ decreases and $\theta$ increases. Photo-alignment dominates after this transition. The sheet shrinks along $\boldsymbol{e}_1$ due to the rotation of nematic director.

Fig. 8 also shows the emergence of a nonzero shear strain $\lambda_{31}$ during the mechano- to photo-alignment transition, while the shear strain $\lambda_{13}$ remains zero. Following Section 3.2, we can again construct stripe domains in the material during this transition. Note in (5.2) that the total free energy only depends on the quadratic of $\lambda_{31}$. As a result, one can find two equilibrium states of homogeneous deformation, denoted as $\boldsymbol{F}^+$ and $\boldsymbol{F}^-$, with $\pm\lambda_{31}$ and $\pm\theta$ respectively, where $\theta \in (0, \pi/2)$. The average deformation gradient is $\langle \boldsymbol{F} \rangle = (\boldsymbol{F}^+ + \boldsymbol{F}^-)/2 = \mathrm{diag}(\lambda_{11}, \lambda_{22}, \lambda_{33})$, such that macroscopically the material is under no shear strain. The normal direction of the domain interface is along $\boldsymbol{e}_1$, and the compatibility condition, $\boldsymbol{F}^+ - \boldsymbol{F}^- = \boldsymbol{a} \otimes \boldsymbol{N}$, is satisfied with $\boldsymbol{a} = (0, 0, 2\lambda_{31})^T$ and $\boldsymbol{N} = \boldsymbol{e}_1$.

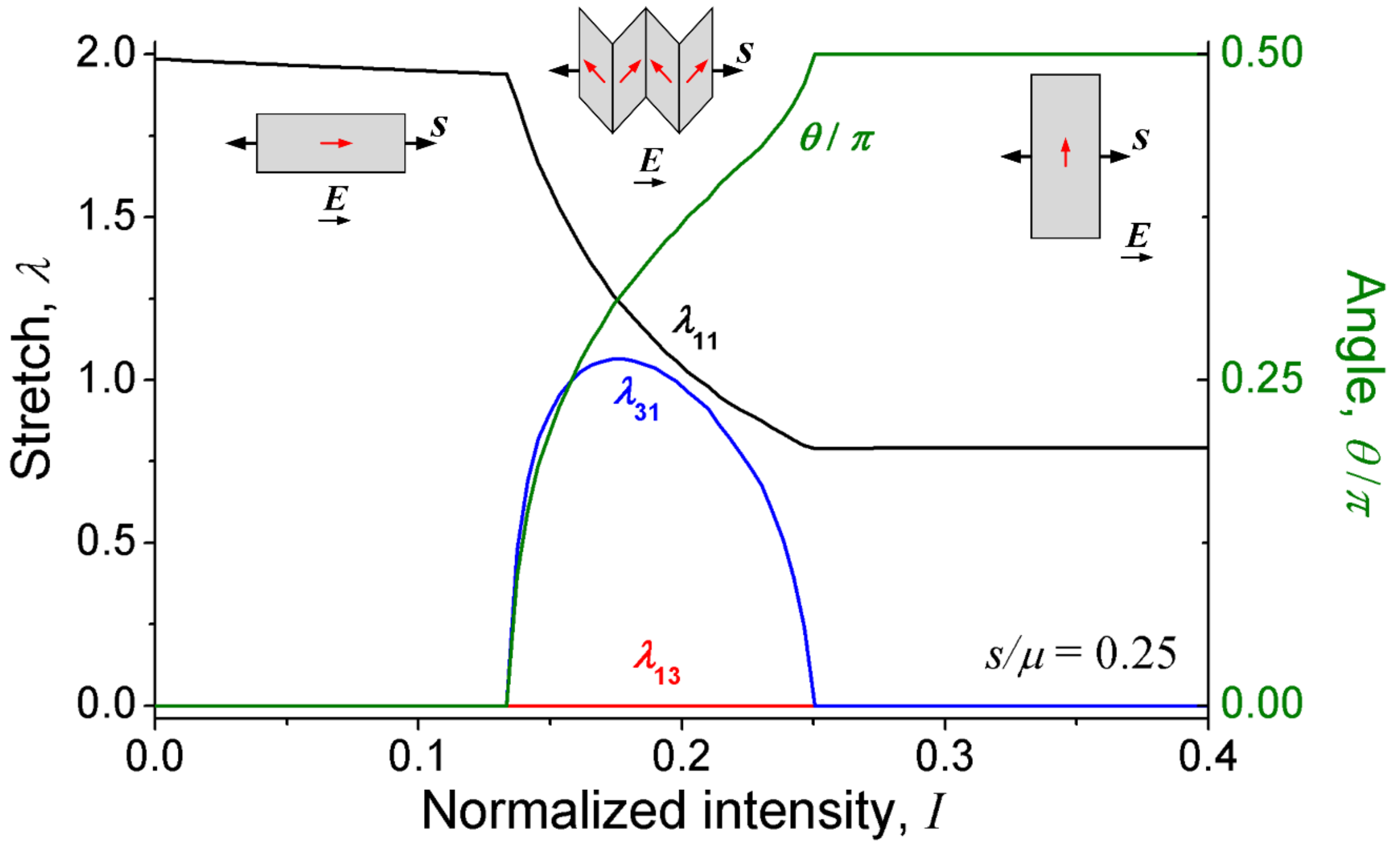

**Fig. 8.** The evolution of stretches and nematic director angle $\theta$ with the increasing light intensity $I$. The nominal stress is uniaxial along $\boldsymbol{e}_1$ and fixed to be $s/\mu = 0.25$. Inset schematics shows the deformation of the material in the regions of mechano-alignment, transition, and photo-alignment. The nematic director is represented by the red arrows. The other branch of solutions of $\lambda_{31}$ and $\theta$ with the same magnitudes but negative sign is not plotted here.

When the light intensity is fixed and the stress increases from zero, we observe similar transition from photo-alignment at small stress to mechano-alignment at large stress, as shown in Fig. 9. Again, stripe domains can form during the transition.

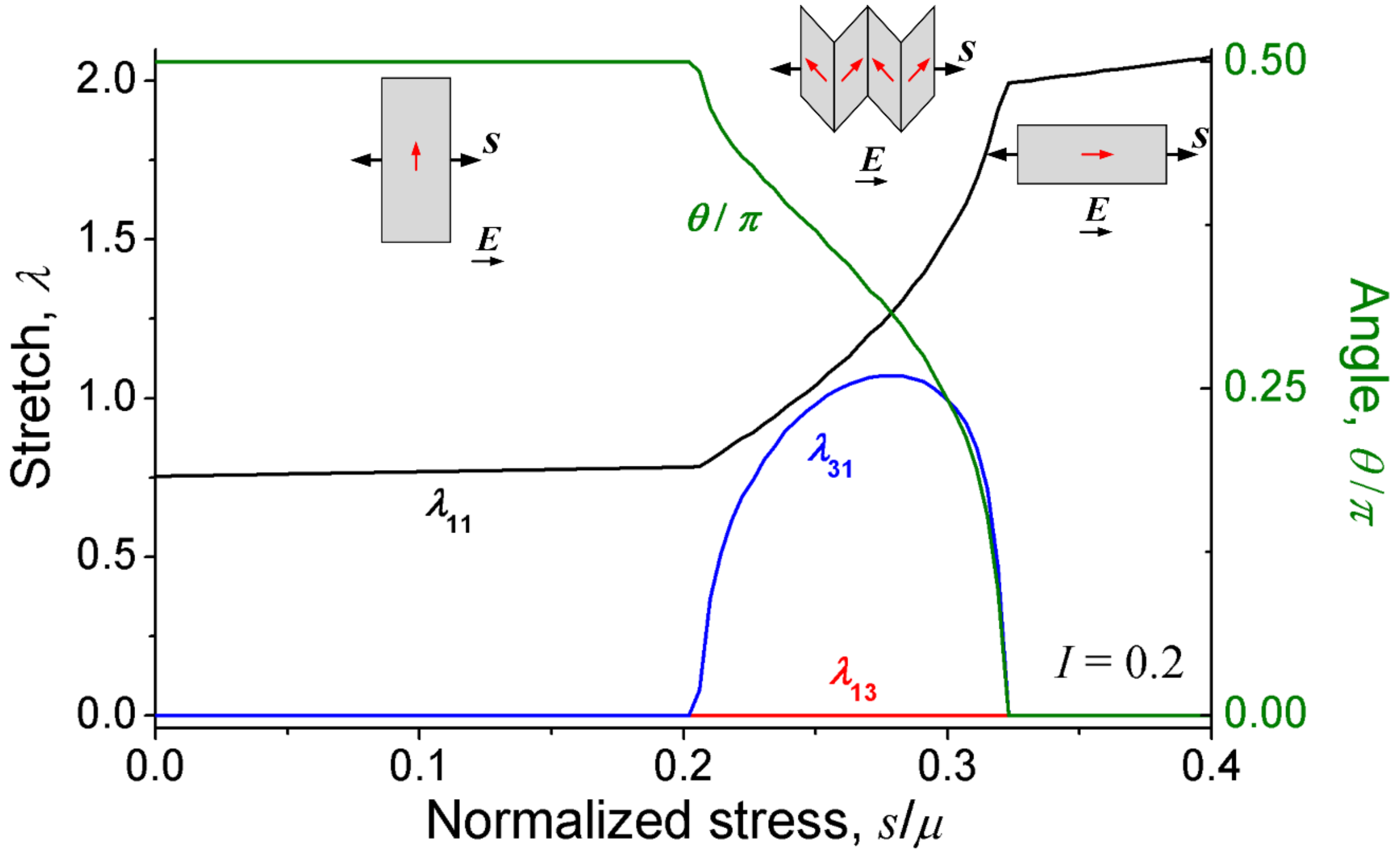

**Fig. 9.** The deformation induced by photo-alignment at small stress and by mechano-alignment at large stress, under a prescribed light intensity $I$ = 0.2. The nematic director is indicated by the angle $\theta$. Inset schematics show the deformation of the material and the stripe domains during the transition region, where the nematic director is represented by the red arrows. The other branch of solutions of $\lambda_{31}$ and $\theta$ with the same magnitudes but negative sign is not plotted here.

Fig. 10 plots an extension of the stretch curves in Fig. 8 under the same set of parameters. After the complete transition to photo-alignment, a further increase of light intensity ultimately induces the first-order, nematic-isotropic phase transformation in the material, which has also been shown previously in Fig. 5c in the case of no stress. The deformation of LCE under increasing light intensity can thus be divided into four regimes in Fig. 10: mechano-alignment at the beginning, a transition towards photo-alignment (the regime of nonzero shear strain), photo-alignment, and the final isotropic phase. Correspondingly, the uniaxial stretch $\lambda_{11}$ is large at first, decreasing in the transition, slowly increasing due to photo-isomerization, and finally increasing sharply due to the first-order phase transformation.

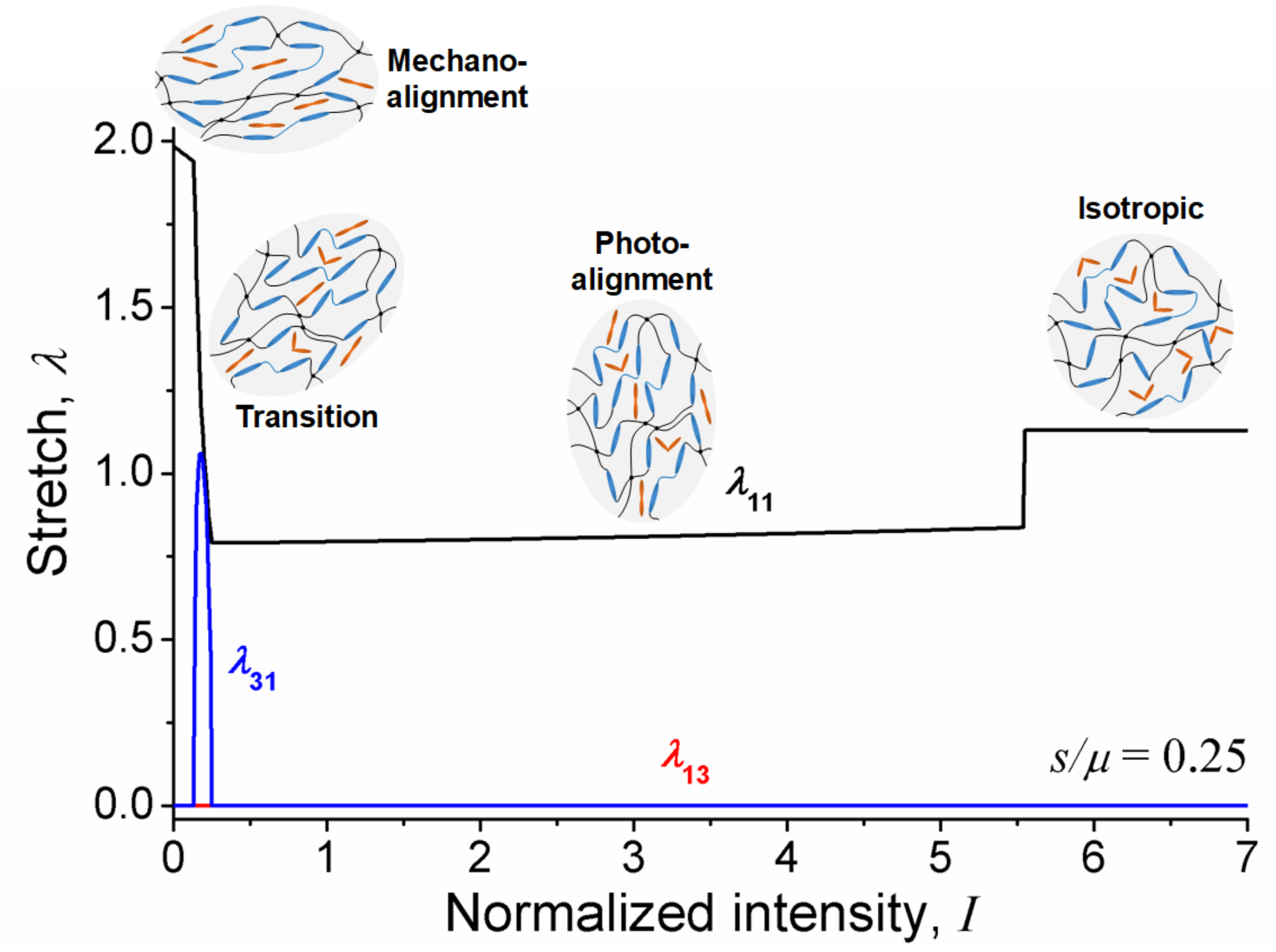

**Fig. 10.** An extension of the stretch curves in Fig. 8 under the same set of parameters. Four regimes of deformation are identified with their corresponding molecular pictures: mechano-alignment, transition, photo-alignment, and isotropic. The shear strain $\lambda_{13}$ remains zero in the entire range of $I$.

When the fixed stress $s$ is large enough, e.g., $s/\mu = 1$ as in Fig. 11, mechano-alignment always dominates even as the light intensity increases. Eventually, when the light intensity is very large as in Fig. 11c, the nematic-isotropic phase transformation is suppressed by the stress. Such observation provides the evidence of a critical point in the phase transformation with the increasing mechanical stress. The first-order transformation takes place below this critical point, but vanishes above it. In the following section, we explore this critical point using the current model.

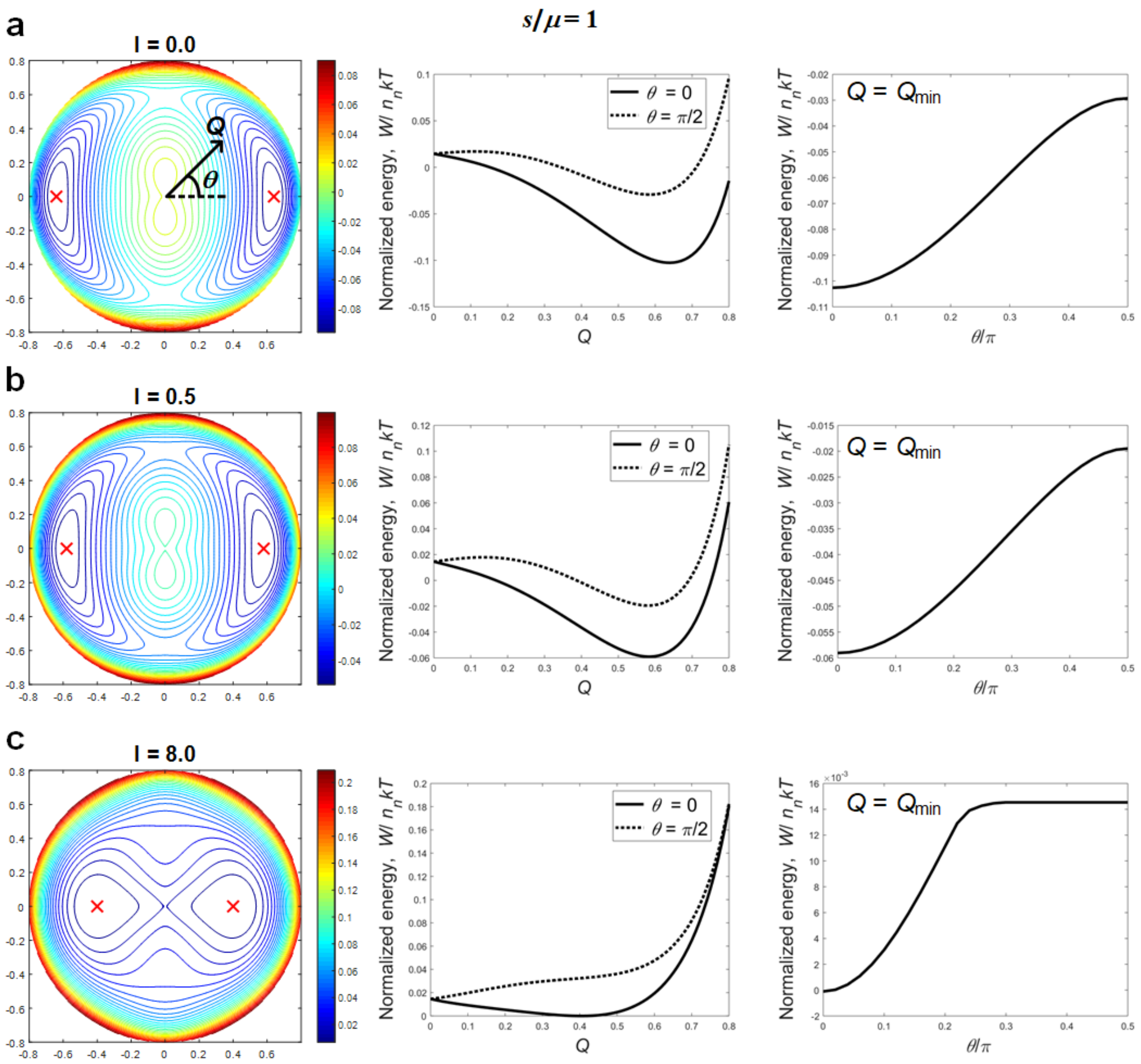

**Fig. 11**. The total energy contour in the polar coordinates $(Q,\theta)$ under a large stress $s/\mu = 1$ and increasing light intensity $I$. The red cross marks the global energy minimum. The nematic director always aligns along the uniaxial stress $s$, indicated by the global energy minimum. In addition, the global minimum at a finite $Q$ in **(c)** shows that when the light intensity is very large, the large stress still suppresses the nematic-isotropic phase transformation.

In general, when a nematic LCE is subject to nonzero mechanical stress or stretch, the field of nematic director at equilibrium depends on the state of the mechanical load. For example, a uniaxial stretch tends to align the director along the stretching direction. The rotation of the director towards the stretching direction induces the soft elastic behavior. As another example, when a nematic LCE sheet is subject to a state of biaxial stretch, depending on the stretch magnitudes, the deformation of the sheet shows phases of liquid-like, wrinkling, microstructure, or solid-like (Cesana et al., 2015; DeSimone and Dolzmann, 2002). To our best knowledge, we are not aware of any experimental study on the coupling between photo- and mechano-alignments discussed in this section.

## 6. First-order phase transformation, snap-through, and critical point

To further explore the first-order phase transformation and critical point, we now consider a case where the polarization and stress are perpendicular to each other, such that both photo- and mechano-alignments align the nematic director along the uniaxial stress (Fig. 12a). The deformation gradient is then

$$\boldsymbol{F} = \begin{bmatrix} \lambda & & \\ & 1/\sqrt{\lambda} & \\ & & 1/\sqrt{\lambda} \end{bmatrix}. \tag{6.1}$$

The balance of force is simplified as

$$\frac{\lambda}{1+2Q} - \frac{1}{(1-Q)\lambda^2} = s/\mu . \tag{6.2}$$

We now fix the stress $s$ at different magnitudes, plot the light intensity $I$ as the loading parameter in the vertical axis, and plot the stretch $\lambda$ and order parameter $Q$ in the horizontal axis in Fig. 12b and 12c, respectively. The increase of light intensity $I$ induces a reduction of order parameter $Q$. Further, note that for small enough stress ($s/\mu$ = 0.5), multiple equilibrium states are present at a fixed light intensity, indicating first-order phase transformation and hysteresis. When the stress is large ($s/\mu$ = 0.75 and 1), this first-order phase transformation is suppressed, and the light intensity approaches infinity at a finite $Q$.

The first-order phase transformation leads to a snap-through instability of the stretch $\lambda$ (arrows in Fig. 12c blue curve). As $I$ increases from 0, the reduction of order parameter decreases the tensile stretch $\lambda$. Further increase of $I$ beyond the peak of curve causes $\lambda$ to immediately snap to a much smaller value. Similarly, the decrease of $I$ from a large value will lead to a backward snap. When the prescribed stress is large enough, no snap-through is observed as the phase transformation is suppressed (Fig. 12c black).

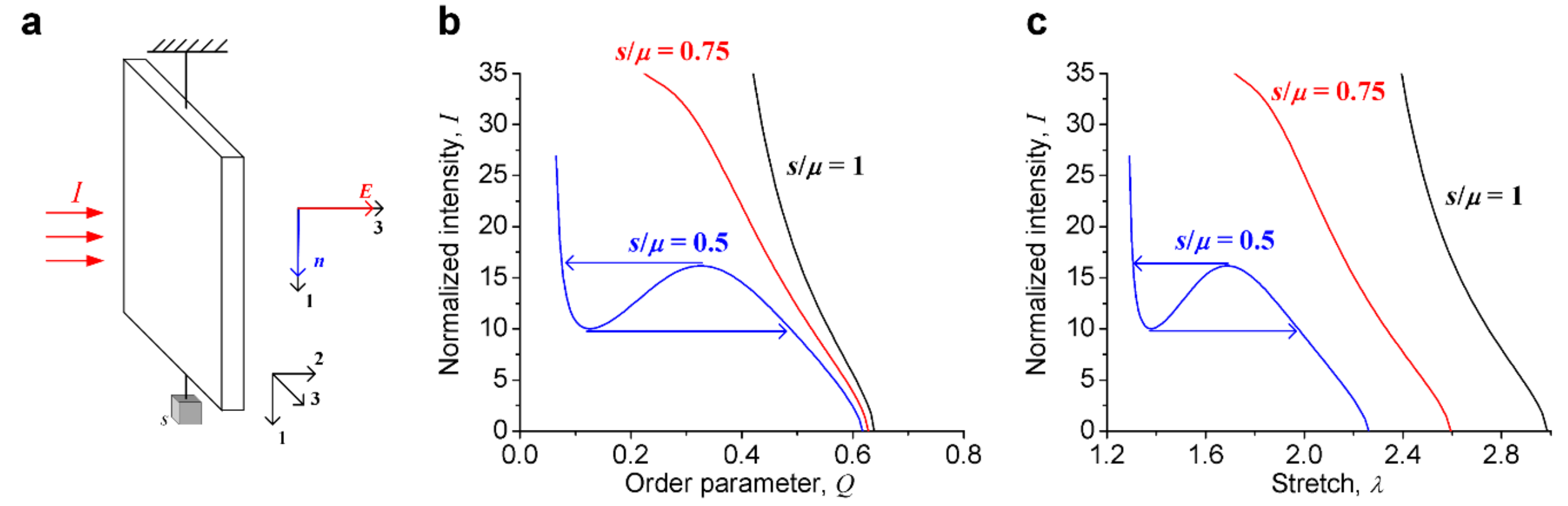

**Fig. 12**. **(a)** In a 2D nematic sheet, both the photo- and mechano-alignments align the nematic director along the uniaxial stress. **(b)** The normalized intensity as a function of the order parameter under different stress levels. When the stress is small ($s/\mu = 0.5$), first-order phase transformation is observed, indicated by the coexisting equilibrium states and hysteresis. When the stress is large ($s/\mu = 0.75$ and 1), the phase transformation is suppressed. **(c)** The normalized intensity as a function of the stretch. When the stress is small ($s/\mu = 0.5$), snap-through instability is observed. When the stress is large ($s/\mu = 0.75$ and 1), no instability is observed.

The disappearance of hysteresis with the increasing stress indicates a transition from a subcritical behavior to a supercritical behavior. To more closely view this transition and the critical point, we numerically plot the phase diagram of the order parameter $Q$, under different light intensity and uniaxial stress (Fig. 13a). We also separately calculate the critical point by letting $\left(\partial I / \partial Q\right)\big|_s = \left(\partial^2 I / \partial Q^2\right)\big|_s = 0$ in Fig. 12b, and mark it on the phase diagram (red dots in Fig. 13). As expected, Below the critical point, a phase boundary is shown between the isotropic and nematic phases, indicated by the discontinuity of the order parameter $Q$. Above the critical point, the first-order phase transformation vanishes, and the equilibrium order parameter changes continuously.

We also plot the same phase diagram for the case when the light polarization is parallel to the uniaxial stress (Fig. 13b). To do so, we neglect the competition between photo- and mechano-alignments, assuming that the nematic director always aligns with the uniaxial stress. The critical point in this case is located at a much lower light intensity, as the coupling between the light and mesogens is much stronger compared to the perpendicular case.

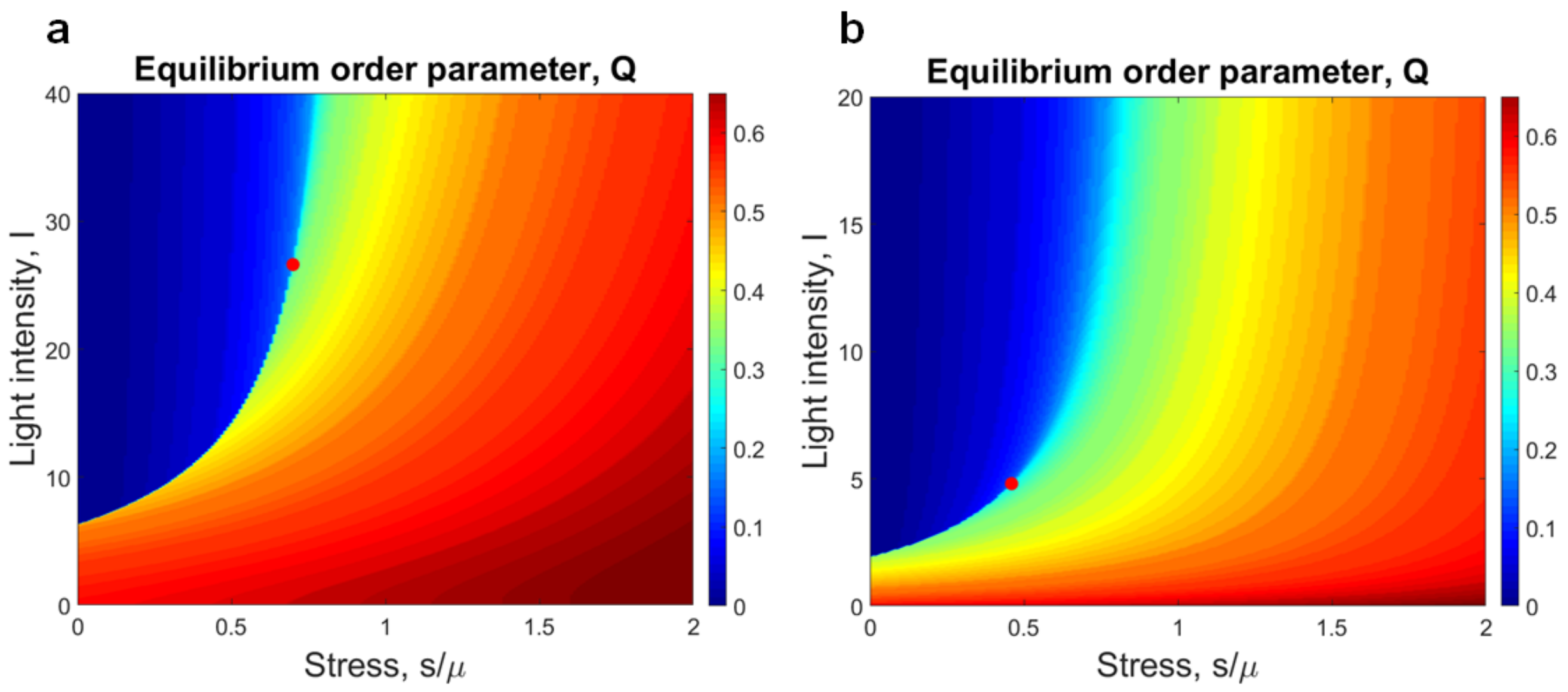

**Fig. 13.** Photomechanical phase diagram and critical point. The red dot marks the numerically calculated critical point. Below the critical point, a phase boundary is shown between the isotropic and nematic phases, indicated by the discontinuity of the equilibrium order parameter $Q$. Above the critical point, the first-order phase transformation vanishes. **(a)** The light polarization is perpendicular to the uniaxial stress $s$. **(b)** The light polarization is parallel to the uniaxial stress $s$.

Light-induced reduction of nematic order parameter and subsequent nematic-isotropic transition has been reported experimentally in their original paper by Finkelmann et al. (2001). However, the effect of mechanical stress on this phase transition has not been well studied. A similar dependence of nematic-isotropic transition on stress was characterized and analyzed in thermomechanical LCEs under varying temperature and stress (Disch et al., 1994; Jin et al., 2010b; Kaufhold et al., 1991; Schätzle et al., 1989). In these experiments with increasing temperature, a crosslinked LCE often shows a continuous phase transformation rather than a first-order discontinuity. Factors that may lead to this phenomenon include material heterogeneity (Selinger et al., 2002) and residual stress in the polymer network (Lebar et al., 2005). The effect of these factors on the light-induced transformation in a LCE deserves further investigation.

## 7. Conclusion

We have theoretically explored photomechanical coupling in a photoactive LCE under both light illumination and mechanical stress. With a linearly polarized light and uniaxial stress, the nematic director in a LCE aligns perpendicular to the light polarization, but parallel to the stress direction. We demonstrate the induced large deformation and the formation of stripe domains during the transition between the two alignments. When the light intensity increases further, a first-order nematic-isotropic phase transformation is observed, with a corresponding snap-through instability in the deformation of the material. A critical point of this phase transformation is identified, above which the order parameter changes continuously.

In this paper, we study photomechanical coupling in a thin sheet of LCE, and assume no light decay through the thickness. In experiment, this scenario can be approximated when the concentration of photochromophores is low enough, or the thickness of the sheet is comparable to the decay length of light in the material. This scenario can be further achieved if the LCE is embedded with *negative photochromophores*, which allow more photons to penetrate upon photoisomerization. When the thickness of sheet is large compared to the decay length of light, the light will only actuate a thin layer of material near the surface, and often lead to a bending actuation. For linearly polarized light, this bending actuation depends on the polarization, as has been shown in both experiment (Yu et al., 2003) and theoretical modeling (Dunn, 2007). Nevertheless, for an arbitrary geometry of LCE under both light illumination and mechanical stress, the theoretical study in the current paper assuming no light decay still provides a simple way of understanding the piecewise local response of the material.

Photo-alignment, mechano-alignment, and light induced nematic-isotropic phase transition have been experimentally studied in separate cases. However, the photomechanical coupling in a LCE under both light illumination and mechanical stress is largely unexplored in experiments. We hope the theoretical results in the current paper will motivate further experimental validation, and help the development of new actuation modes in LCEs. The results are also hoped to motivate future studies on photomechanical coupling with more complex loading conditions, as well as kinetic models for photomechanical actuation.

**Acknowledgements**

This work was supported by the Office of Naval Research through the MURI on Photomechanical Material Systems (ONR N00014-18-1-2624). We acknowledge the fruitful discussion with M. Ravi Shankar on the photomechanical phase diagram.

**Appendix A. Derivation of the Maier-Saupe mean field model**

The derivation of the Maier-Saupe mean field model (Maier and Saupe, 1959) follows the work by Corbett and Warner (2006, 2008). Consider a closed system with $n$ mesogens. The total Hamiltonian of this system is assumed to be induced by the dispersion force between individual mesogens, expressed as

$$H = -\frac{1}{2}\sum_{i \neq j}^{n} J_{ij} Q_i Q_j \, , \tag{A1}$$

where the sum takes over all the mesogens, $J_{ij}$ denotes the interaction potential between the $i_{\text{th}}$ and $j_{\text{th}}$ mesogens, $Q_i = \left(3\cos^2\alpha_i - 1\right)/2$, and $\alpha_i$ is the angle between the long axis of the $i_{\text{th}}$ mesogen and the nematic director.

To formulate a mean field model, we choose a trial field with potential $h$, so that the trial Hamiltonian of the system under this trial field is

$$H_0 = -h\sum_{i}^{n} Q_i \, , \tag{A2}$$

where the trial Hamiltonian of each mesogen is $-hQ_i$. Because all the mesogens are statistically independent of each other under this trial field, the partition function based on the canonical ensemble under the trial field is

$$Z_{n,0} = \prod_{i}^{n} \int_{-1}^{1} 2\pi e^{hQ_i/kT} \mathrm{d}\left(\cos\alpha_i\right) = \left(4\pi\right)^n Z_0^n \, , \tag{A3}$$

where $Z_0 = \int_0^1 e^{hQ_i/kT} \mathrm{d}\left(\cos\alpha_i\right)$, and we have simplified (A3) based on the symmetry of the integration. The change of Helmholtz free energy due to the trial field is

$$F_0 = -nkT\log Z_0 \, , \tag{A4}$$

in which we have subtracted the Helmholtz free energy $-nkT\log\left(4\pi\right)$ of the initial system where there is no external field or any interaction.

To obtain the Helmholtz free energy of the system based on the Hamiltonian in (A1), we apply the

Bogolyubov inequality (Feynman et al., 1972)

$$F \le F_0 + \langle H - H_0 \rangle_0 , \tag{A5}$$

where $\langle \ \rangle_0$ denotes the statistical mean based on the canonical ensemble under the trial field Hamiltonian $H_0$. We define the nematic order parameter $Q$ as

$$Q = \langle Q_i \rangle_0 = \frac{\int_0^1 Q_i e^{hQ_i/kT} \mathrm{d}(\cos\alpha_i)}{\int_0^1 e^{hQ_i/kT} \mathrm{d}(\cos\alpha_i)} . \tag{A6}$$

Equivalently,

$$Q = kT \frac{\mathrm{d}(\log Z_0)}{\mathrm{d}\lambda} = g\left(\frac{h}{kT}\right). \tag{A7}$$

Under the trial field, the mesogens are statistically independent of each other, so we can simply calculate the mean of (A1) and (A2) to get

$$\langle H - H_0 \rangle_0 = -\frac{1}{2} nJQ^2 + nhQ , \tag{A8}$$

where $J = \frac{1}{n} \sum_{i \ne j}^{n} J_{ij}$ is the average interaction potential. We express the Helmholtz free energy $F$ using the upper bound on the right of (A5):

$$F(Q) = n\left(-kT \log Z_0 + kTg^{-1}(Q)Q - \frac{1}{2}JQ^2\right), \tag{A9}$$

where we have used $h = kTg^{-1}(Q)$ from (A7). We can further express

$$g(x) = -\frac{1}{2} - \frac{1}{2x} + \frac{1}{2x}\sqrt{\frac{3x}{2}} \frac{\exp(3x/2)}{\int_0^{\sqrt{3x/2}} \exp(y^2)\mathrm{d}y} , \tag{A10}$$

and

$$Z_0 = \int_0^1 e^{g^{-1}(Q)Q_i} \mathrm{d}(\cos\alpha_i) = \frac{\exp\left[g^{-1}(Q)\right]}{1 + g^{-1}(Q)(1+2Q)} . \tag{A11}$$

**Appendix B. Calculation of the *cis*-fraction in the total mesogens**

Following Corbett and Warner (2006, 2008), we calculate $c$ as the fraction of the *cis*-mesogens in the total mesogens $N_n$. During the isomerization, the sum of the *trans*-mesogens $N_t$ and the *cis*-mesogens $N_c$ is conserved, and is prescribed as a constant fraction $f$ of the total mesogens $N_n$:

$$N_t + N_c = fN_n, \tag{B1}$$

In the equilibrium state, consider a single mesogen with its long axis in direction $\boldsymbol{u}$, forming an angle of $\alpha_u$ with the nematic director $\boldsymbol{n}$ (Fig. 1b). The nematic order parameter $Q$ is defined as

$$Q = \left\langle \frac{1}{2}\left(3\cos^2\alpha_u - 1\right) \right\rangle, \tag{B2}$$

where $\langle\ \rangle$ calculates the mean value of the term inside based on the statistics of all the *trans*-mesogens and non-photoactive mesogens.

The forward reaction rate of the isomerization is proportional to the product of the mean projected light intensity on $\boldsymbol{u}$ over all the *trans*-mesogens, $\left\langle (\boldsymbol{E}\cdot\boldsymbol{u})^2 \right\rangle_{trans\text{-}}$, and the concentration of the *trans*-mesogens $N_t$:

$$r_{trans\text{-}cis} = \Gamma \left\langle (\boldsymbol{E}\cdot\boldsymbol{u})^2 \right\rangle_{trans\text{-}} N_t, \tag{B3}$$

where $\Gamma$ is a constant, and $\boldsymbol{E}$ is the electric field indicating the polarization of the light. Because we assume the *trans*- and the non-photoactive mesogens follow the same statistics, we have

$$\left\langle (\boldsymbol{E}\cdot\boldsymbol{u})^2 \right\rangle_{trans\text{-}} = \left\langle (\boldsymbol{E}\cdot\boldsymbol{u})^2 \right\rangle. \tag{B4}$$

The thermally induced backward reaction rate is proportional to the concentration of the *cis*-mesogens $N_c$:

$$r_{cis\text{-}trans} = N_c / \tau, \tag{B5}$$

where $\tau$ is a constant indicating the relaxation time of the *cis*-state. In equilibrium, the forward and backward reaction rates are equal. Equating (B3) and (B5), using the conservation of mass (B1), and recalling that $c = N_c/N_n$, we obtain

$$c = \frac{f\Gamma\tau\left\langle\left(\boldsymbol{E}\cdot\boldsymbol{u}\right)^2\right\rangle}{1+\Gamma\tau\left\langle\left(\boldsymbol{E}\cdot\boldsymbol{u}\right)^2\right\rangle}. \tag{B6}$$

Now consider the case shown in Fig. 1b on the right, where the three vectors $\boldsymbol{u}$, $\boldsymbol{n}$, and $\boldsymbol{E}$ form a vertex of a tetrahedron. The angles between the three vectors are denoted as $<\boldsymbol{u},\boldsymbol{n}>=\alpha_u$, $<\boldsymbol{n},\boldsymbol{E}>=\theta$, and $<\boldsymbol{u},\boldsymbol{E}>$. In addition, we denote $\varsigma_n$ as the dihedral angle between the surface formed by $\left(\boldsymbol{u},\boldsymbol{n}\right)$ and the surface formed by $\left(\boldsymbol{n},\boldsymbol{E}\right)$.

The average projected light intensity over all the mesogens is calculated as

$$\left\langle\left(\boldsymbol{E}\cdot\boldsymbol{u}\right)^2\right\rangle = \boldsymbol{E}^2\left\langle\cos^2<\boldsymbol{u},\boldsymbol{E}>\right\rangle, \tag{B7}$$

The trigonometric relation in a tetrahedron gives that

$$\cos<\boldsymbol{u},\boldsymbol{E}>=\cos\alpha_u\cos\theta+\sin\alpha_u\sin\theta\cos\varsigma_n. \tag{B8}$$

In addition, the definition of nematic order parameter in (B2) gives

$$\left\langle\cos^2\alpha_u\right\rangle = \frac{2Q+1}{3}. \tag{B9}$$

The axisymmetric feature of the distribution of $\boldsymbol{u}$ around the nematic director $\boldsymbol{n}$ further gives

$$\left\langle\cos\varsigma_n\right\rangle = 0, \tag{B10}$$

$$\left\langle\cos^2\varsigma_n\right\rangle = \frac{1}{2}. \tag{B11}$$

Substituting (B8) – (B11) into (B7), and again using the axisymmetry of the distribution of $\boldsymbol{u}$, we obtain

$$\left\langle\left(\boldsymbol{E}\cdot\boldsymbol{u}\right)^2\right\rangle = \frac{1}{3}\boldsymbol{E}^2\left[1+Q\left(3\cos^2\theta-1\right)\right]. \tag{B12}$$

As a result,

$$c = c\left(Q,\theta\right) = f\frac{I\left[1+Q\left(3\cos^2\theta-1\right)\right]}{3+I\left[1+Q\left(3\cos^2\theta-1\right)\right]}, \tag{B13}$$

where $I=\boldsymbol{E}^2\Gamma\tau$ is the dimensionless intensity of the light.

## Reference

Ahn, S.-k., Ware, T.H., Lee, K.M., Tondiglia, V.P., White, T.J., 2016. Photoinduced Topographical Feature Development in Blueprinted Azobenzene-Functionalized Liquid Crystalline Elastomers. Adv. Funct. Mater. 26, 5819-5826.

Bai, R., Suo, Z., 2015. Optomechanics of Soft Materials. J. Appl. Mech. 82, 071011-071011-071019.

Bhattacharya, K., 2003. Microstructure of martensite: why it forms and how it gives rise to the shape-memory effect. Oxford University Press.

Biggins, J.S., Warner, M., Bhattacharya, K., 2009. Supersoft elasticity in polydomain nematic elastomers. Phys. Rev. Lett. 103, 037802.

Biggins, J.S., Warner, M., Bhattacharya, K., 2012. Elasticity of polydomain liquid crystal elastomers. J. Mech. Phys. Solids 60, 573-590.

Bladon, P., Terentjev, E.M., Warner, M., 1993. Transitions and instabilities in liquid crystal elastomers. Phys. Rev. E 47, R3838.

Camacho-Lopez, M., Finkelmann, H., Palffy-Muhoray, P., Shelley, M., 2004. Fast liquid-crystal elastomer swims into the dark. Nat. Mater. 3, 307.

Cesana, P., Plucinsky, P., Bhattacharya, K., 2015. Effective behavior of nematic elastomer membranes. Archive for Rational Mechanics and Analysis 218, 863-905.

Corbett, D., Warner, M., 2006. Nonlinear Photoresponse of Disordered Elastomers. Phys. Rev. Lett. 96, 237802.

Corbett, D., Warner, M., 2007. Linear and nonlinear photoinduced deformations of cantilevers. Phys. Rev. Lett. 99, 174302.

Corbett, D., Warner, M., 2008. Polarization dependence of optically driven polydomain elastomer mechanics. Phys. Rev. E 78, 061701.

Cviklinski, J., Tajbakhsh, A.R., Terentjev, E.M., 2002. UV isomerisation in nematic elastomers as a route to photo-mechanical transducer. The European Physical Journal E 9, 427-434.

de Gennes, P.G., Prost, J., 1995. The Physics of Liquid Crystals. Clarendon Press.

DeSimone, A., Dolzmann, G., 2002. Macroscopic Response of¶ Nematic Elastomers via Relaxation of¶ a Class of SO (3)-Invariant Energies. Archive for rational mechanics and analysis 161, 181-204.

Disch, S., Schmidt, C., Finkelmann, H., 1994. Nematic elastomers beyond the critical point. Macromol. Rapid Commun. 15, 303-310.

Dong, X., Tong, F., Hanson, K.M., Al-Kaysi, R.O., Kitagawa, D., Kobatake, S., Bardeen, C.J., 2019. Hybrid Organic–Inorganic Photon-Powered Actuators Based on Aligned Diarylethene Nanocrystals. Chem. Mater. 31, 1016-1022.

Dunn, M.L., 2007. Photomechanics of mono-and polydomain liquid crystal elastomer films. J. Appl. Phys. 102, 013506.

Feynman, R.P., Kikuchi, R., Shaham, J., Feiveson, H.A., Shaham, M., 1972. Statistical Mechanics: A Set of Lectures. W. A. Benjamin.

Finkelmann, H., Nishikawa, E., Pereira, G.G., Warner, M., 2001. A new opto-mechanical effect in solids. Phys. Rev. Lett. 87, 015501.

Frank, F.C., 1958. I. Liquid crystals. On the theory of liquid crystals. Discuss. Faraday Soc. 25, 19-28.

Gelebart, A.H., Mulder, D.J., Varga, M., Konya, A., Vantomme, G., Meijer, E.W., Selinger, R.L.B., Broer, D.J., 2017. Making waves in a photoactive polymer film. Nature 546, 632.

Hogan, P.M., Tajbakhsh, A.R., Terentjev, E.M., 2002. UV manipulation of order and macroscopic shape in nematic elastomers. Phys. Rev. E 65, 041720.

Ichimura, K., 2000. Photoalignment of liquid-crystal systems. Chem. Rev. 100, 1847-1874.

Ikeda, T., 2003. Photomodulation of liquid crystal orientations for photonic applications. J. Mater. Chem. 13, 2037-2057.

Jin, L., Yan, Y., Huo, Y., 2010a. A gradient model of light-induced bending in photochromic liquid crystal elastomer and its nonlinear behaviors. International Journal of Non-Linear Mechanics 45, 370-381.

Jin, L., Zeng, Z., Huo, Y., 2010b. Thermomechanical modeling of the thermo-order–mechanical coupling behaviors in liquid crystal elastomers. J. Mech. Phys. Solids 58, 1907-1927.
Kaufhold, W., Finkelmann, H., Brand, H.R., 1991. Nematic elastomers, 1. Effect of the spacer length on the mechanical coupling between network anisotropy and nematic order. Die Makromolekulare Chemie: Macromolecular Chemistry and Physics 192, 2555-2579.
Kim, T., Zhu, L., Al-Kaysi, R.O., Bardeen, C.J., 2014. Organic Photomechanical Materials. Chemphyschem 15, 400-414.
Knežević, M., Warner, M., Čopič, M., Sánchez-Ferrer, A., 2013. Photodynamics of stress in clamped nematic elastomers. Phys. Rev. E 87, 062503.
Kuenstler, A.S., Hayward, R.C., 2019. Light-induced shape morphing of thin films. Current Opinion in Colloid & Interface Science 40, 70-86.
Küupfer, J., Finkelmann, H., 1994. Liquid crystal elastomers: Influence of the orientational distribution of the crosslinks on the phase behaviour and reorientation processes. Macromol. Chem. Phys. 195, 1353-1367.
Lebar, A., Kutnjak, Z., Žumer, S., Finkelmann, H., Sánchez-Ferrer, A., Zalar, B., 2005. Evidence of supercritical behavior in liquid single crystal elastomers. Phys. Rev. Lett. 94, 197801.
Lin, Y., Jin, L., Huo, Y., 2012. Quasi-soft opto-mechanical behavior of photochromic liquid crystal elastomer: Linearized stress–strain relations and finite element simulations. Int. J. Solids Struct. 49, 2668-2680.
Liu, L., Onck, P.R., 2017. Enhanced deformation of azobenzene-modified liquid crystal polymers under dual wavelength exposure: A photophysical model. Phys. Rev. Lett. 119, 057801.
Mahimwalla, Z., Yager, K.G., Mamiya, J.-i., Shishido, A., Priimagi, A., Barrett, C.J., 2012. Azobenzene photomechanics: prospects and potential applications. Polym. Bull. 69, 967-1006.
Maier, W., Saupe, A., 1959. Eine einfache molekular-statistische Theorie der nematischen kristallinflüssigen Phase. Teil l1. Zeitschrift für Naturforschung A 14, 882-889.
Olmsted, P.D., 1994. Rotational invariance and Goldstone modes in nematic elastomers and gels. J. Phys. II 4, 2215-2230.
Pang, X., Lv, J.a., Zhu, C., Qin, L., Yu, Y., 2019. Photodeformable Azobenzene-Containing Liquid Crystal Polymers and Soft Actuators. Adv. Mater.
Schätzle, J., Kaufhold, W., Finkelmann, H., 1989. Nematic elastomers: The influence of external mechanical stress on the liquid-crystalline phase behavior. Die Makromolekulare Chemie: Macromolecular Chemistry and Physics 190, 3269-3284.
Selinger, J.V., Jeon, H.G., Ratna, B.R., 2002. Isotropic-Nematic Transition in Liquid-Crystalline Elastomers. Phys. Rev. Lett. 89, 225701.
Ube, T., Ikeda, T., 2014. Photomobile Polymer Materials with Crosslinked Liquid-Crystalline Structures: Molecular Design, Fabrication, and Functions. Angew. Chem. Int. Ed. 53, 10290-10299.
Urayama, K., Kohmon, E., Kojima, M., Takigawa, T., 2009. Polydomain−Monodomain Transition of Randomly Disordered Nematic Elastomers with Different Cross-Linking Histories. Macromolecules 42, 4084-4089.
Verwey, G.C., Warner, M., Terentjev, E.M., 1996. Elastic instability and stripe domains in liquid crystalline elastomers. J. Phys. II 6, 1273-1290.
Warner, M., Bladon, P., Terentjev, E.M., 1994. "Soft elasticity"—deformation without resistance in liquid crystal elastomers. J. Phys. II 4, 93-102.
Warner, M., Mahadevan, L., 2004. Photoinduced Deformations of Beams, Plates, and Films. Phys. Rev. Lett. 92, 134302.
Warner, M., Terentjev, E.M., 2003. Liquid Crystal Elastomers. OUP Oxford.
White, T.J., 2018. Photomechanical effects in liquid crystalline polymer networks and elastomers. J. Polym. Sci., Part B: Polym. Phys. 56, 695-705.
White, T.J., Tabiryan, N.V., Serak, S.V., Hrozhyk, U.A., Tondiglia, V.P., Koerner, H., Vaia, R.A., Bunning, T.J., 2008. A high frequency photodriven polymer oscillator. Soft Matter 4, 1796-1798.
Wie, J.J., Shankar, M.R., White, T.J., 2016. Photomotility of polymers. Nat. Commun. 7, 13260.

Yamada, M., Kondo, M., Mamiya, J.-i., Yu, Y., Kinoshita, M., Barrett, C.J., Ikeda, T., 2008. Photomobile Polymer Materials: Towards Light-Driven Plastic Motors. Angew. Chem. Int. Ed. 47, 4986-4988.
Yu, Y., Nakano, M., Ikeda, T., 2003. Directed bending of a polymer film by light. Nature 425, 145-145.
Zeng, H., Wasylczyk, P., Wiersma, D.S., Priimagi, A., 2018. Light robots: bridging the gap between microrobotics and photomechanics in soft materials. Adv. Mater. 30, 1703554.